\documentclass[useAMS,usenatbib, ]{mn2e}

\usepackage{aas_macros}
\usepackage{amsfonts,amssymb,amsmath}
\usepackage{comment}
\usepackage{graphicx}
\usepackage{graphics}
\usepackage{epstopdf} 
\usepackage[usenames,dvipsnames]{xcolor}

\usepackage{url}

\title[The Elusive Stellar Halo of M33]{The elusive stellar halo of the Triangulum Galaxy\thanks{This work was based on observations obtained with the MegaPrime/MegaCam, a joint project of CFHT and CEA/DAPNIA, at the Canada-France-Hawaii Telescope (CFHT) which is operated by the National Research Council (NRC) of Canada, the Institute National des Sciences de l'Univers of the Centre National de la Recherche Scientifique of France and the University of Hawaii.}}
\author[B. McMonigal et al.]
	{B. McMonigal$^{1}$\thanks{E-mail: b.mcmonigal@physics.usyd.edu.au (BM)}, G. F. Lewis$^{1}$, B. J. Brewer$^{2}$, M. J. Irwin$^{3}$, N. F. Martin$^{4, 5}$, \newauthor A. W. McConnachie$^{4}$, R. A. Ibata$^{4}$, A. M. N. Ferguson$^{7}$, A. D. Mackey$^{8}$, \newauthor S. C. Chapman$^{9}$ \\
$^{1}$Sydney Institute for Astronomy, School of Physics, A28, The University of Sydney, Sydney, NSW 2006, Australia\\
$^{2}$Department of Statistics, The University of Auckland, Private Bag 92019, Auckland 1142, New Zealand\\
$^{3}$Institute of Astronomy, Madingley Road, University of Cambridge, CB3 0HA, UK\\
$^{4}$Observatoire Astronomique, Universite de Strasbourg, CNRS, F-67000 Strasbourg, France\\
$^{5}$Max-Planck-Institut f\"{u}r Astronomie, K\"{o}nigstuhl 17, D-69117 Heidelberg, Germany\\
$^{6}$NRC Herzberg Institute of Astrophysics, 5071 West Saanich Road, Victoria, British Columbia V9E 2E7, Canada\\
$^{7}$Institute for Astronomy, University of Edinburgh, Royal Observatory, Blackford Hill, Edinburgh, EH9 3HJ, UK\\
$^{8}$RSAA, The Australian National University, Mount Stromlo Observatory, Cotter Road, Weston Creek, ACT 2611, Australia\\
$^{9}$Department of Physics and Atmospheric Science, Dalhousie University, Coburg Road, Halifax, NS B3H 1A6, Canada\\}
\begin{document}

\date{Accepted --. Received --; in original form --}

\pagerange{\pageref{firstpage}--\pageref{lastpage}} \pubyear{2016}

\maketitle

\label{firstpage}

\begin{abstract}
The stellar halos of large galaxies represent a vital probe of the processes of galaxy evolution. They are the remnants of the initial bouts of star formation during the collapse of the proto-galactic cloud, coupled with imprint of ancient and on-going accretion events. 
Previously, we have reported the tentative detection of a possible, faint, extended stellar halo in the Local Group spiral, the Triangulum Galaxy (M33). However, the presence of substructure surrounding M33 made interpretation of this feature difficult. 
Here, we employ the final data set from the Pan-Andromeda Archaeological Survey (PAndAS), combined with an improved calibration and a newly derived contamination model for the region to revisit this claim. With an array of new fitting algorithms, fully accounting for contamination and the substantial substructure beyond the prominent stellar disk in M33, we reanalyse the surrounds to separate the signal of the stellar halo and the outer halo substructure. 
Using more robust search algorithms, we do not detect  a large scale smooth stellar halo and place a limit on the maximum surface brightness of such a feature of $\mu_V=35.5$ mags per square arcsec, or a total halo luminosity of $L < 10^6 L_\odot$. \end{abstract}


\section{Introduction}
A key feature of $\Lambda$ Cold Dark Matter ($\Lambda$CDM) cosmological models is the hierarchical formation of structure \citep[see][for an overview]{2010gfe..book.....M}. With this, large galaxies are built up over time through the continual accretion of smaller structures. Accretion progenitors that fall towards the centre of their new host are heavily disrupted and the short dynamical timescales in the inner halo rapidly mix accreted structures to form a smooth stellar background. In the outer halo, where timescales are longer, ongoing accretion events can be found in the form of coherent phase-space stellar streams. However, major mergers can add to the confusion, violently disrupting the host and erasing most information of past accretions. The final resting place of many of the accretion events is the diffuse stellar halo, a faint component making up only a few per cent of the total luminosity of its host galaxy. Hence the properties of these stellar halos represent an archaeological record of the processes that shape a galaxy over cosmic time \citep[e.g.][]{2004MNRAS.349...52B,2005ApJ...635..931B,2015MNRAS.454.3185C}.

Recent focus has turned to studying the stellar halos of Local Group galaxies through the identification of resolved stellar populations, with surveys such as SDSS/SEGUE revealing the extensive halo properties of our own Milky Way \citep[e.g.][]{2015ApJ...809..144X}. The other large galaxies within the Local Group, namely the Andromeda (M31) and Triangulum (M33) galaxies, have been the targets of the Pan-Andromeda Archaeological Survey (PAndAS), uncovering substantial stellar substructure and an extensive halo surrounding Andromeda \citep{Ibata2014}. 

M33 has also been found to possess extensive stellar substructure, in the form of a highly distorted outer disk, thought to have been formed in an interaction with the larger M31 \citep{McConnachie2009,McConnachie2010}; this substructure is roughly aligned with the previously detected distorted HI disk \citep{2009ApJ...703.1486P,2013ApJ...763....4L}. Being about a tenth the size of the two other large galaxies within the Local Group, the properties of any stellar halo of M33 would provide clues to galaxy evolution on a different mass scale than for the Milky Way and M31. 
The smooth halo component around M33 has been extremely elusive; early work presented in \citet{Ibata2007} claimed a detection, which was later shown to be the extended substructure. \citet{Cockcroft2013}, hereafter C13, after excising significant substructure and accounting for foreground contamination, presented a tentative detection of a smooth stellar halo with a scale-length of $\sim$20 kpc, and an estimated total luminosity of a few percent of the luminous disk. 
In this work, we revisit the detection and characterisation of the stellar halo of M33 using new analysis techniques, the final PAndAS data set with improved calibration, and a more detailed contamination model developed from the PAndAS data \citep{Martin2013}. We seek to fully characterise the smooth component of the stellar halo without resorting to masking the lumpy substructure component; ideally this should be recovered and characterised as a byproduct of our analysis. In order to monitor the validity of our results, we thoroughly test our methods using synthetic datasets generated to match the PAndAS data for M33.

Section \ref{sec:data} describes the data and models we employ in investigating the M33 stellar halo. In Section \ref{sec:methods}, we discuss the over-arching methodology we use throughout, including colour-magnitude selection, spatial selection, masking, binning, and most importantly the synthetic data used to test the fitting algorithms. We present the results of our tests in Section \ref{sec:C13}, first of replicating the methods in C13 and then of alternative algorithms, both for the PAndAS data and synthetic mock data. Finally in Section \ref{sec:discussion} we discuss and conclude.

\section{PAndAS Data}\label{sec:data}
The stellar data employed in this study was obtained as part of the Pan-Andromeda Archaeological Survey (PAndAS; \citealt{McConnachie2009}). This Large Program on the 
3.6-metre Canada-France-Hawaii Telescope (CFHT) used the $0.96\times0.94$ square degree field of view MegaPrime camera to map out the halos of M31 and M33 to distances of approximately 150 kpc and 50 kpc, with a total coverage of $\sim390$~deg$^2$. Full details of the data reduction are presented in \citet{Ibata2014}, and a public release of the data is forthcoming (McConnachie et al., in preparation). All observations were taken in good seeing ($ \lesssim 0.8\arcsec$), with a mean seeing of $0.67\arcsec$ in $g$-band and $0.60\arcsec$ in $i$-band. 
The resultant median depth of the survey is $g = 26.0$~mag and $i=24.8$~mag ($5\sigma$). 

The data were pre-processed with CFHT's \textit{Elixir} pipeline, to perform bias, flat, and fringe correction and determine the photometric zero point of the observations.
Further processing was undertaken using a bespoke version of the CASU photometry pipeline \citep{Irwin2001} adapted for CFHT/MegaPrime observations, including re-registration, stacking, catalogue generation and object morphological classification, and creating merged $g$, $i$ catalogues. Based on curve of growth analysis, the pipeline classifies objects as noise detections, galaxies, and probable stars. We employ all objects in the final catalogue that have been reliably classified as stars in both bands (aperture photometry classifications of -1 or -2 in both $g$ and $i$, which corresponds to point sources up to $2\sigma$ from the stellar locus). The CFHT instrumental magnitudes $g$ and $i$ are transformed to de-reddened magnitudes $g_0$ and $i_0$ on a source-by-source basis, using the following relationships from \citet*{Schlegel1998}: $g_0 = g - 3.793E(B-V)$ and $i_0 = i - 2.086E(B-V)$. 

Despite every effort to systematically cover the PAndAS survey region, holes are unavoidable at the location of bright saturated stars, chip gaps, and a few failed CCDs. These holes are filled with fake stars by duplicating information from nearby regions (for details, see \citealt{Ibata2014}). These entries make up only a few percent of the catalogue entries, and are included in the following analysis to best approximate homogeneous coverage of the M33 region. 

\subsection{Contamination Model}\label{sec:contamination}

For the study of the M33 halo, there are three main sources of extensive contamination within the PAndAS footprint: namely foreground faint Milky Way dwarfs and M31 halo giant stars, and at fainter magnitudes unresolved compact background galaxies.
Since the signal of the M33 stellar halo we are searching for is extremely faint, it is critical that the contamination be modelled as accurately as possible. In C13, this contamination was modelled as a constant value, ignoring any spatial variation. This is reasonable for small spatial regions, but is unlikely to be representative of the contamination over a region as large as the M33 stellar halo. More recently, \citet{Martin2013} presented a spatially resolved contamination model developed empirically from the PAndAS data, and it is this model that is employed in this study. With this, the density of contaminants from intervening Milky Way populations,  $\Sigma$, at a given location $(\xi_{M31}, \eta_{M31})$ and a given colour and magnitude $(g_0-i_0, i_0)$ is given by an exponential dependent upon three components:
\begin{multline}
\Sigma_{(g_0-i_0,i_0)}(\xi_{M31}, \eta_{M31}) = \exp(\alpha_{(g_0-i_0,i_0)}\xi_{M31} \\ + \beta_{(g_0-i_0,i_0)}\eta_{M31} +\gamma_{(g_0-i_0,i_0)}).
\end{multline}
The coordinates, $(\xi_{M31},\eta_{M31})$, in this model are a tangent-plane projection centred on M31, although for the remainder of this work we use the coordinates $(\xi,\eta)$ to refer to the tangent plane projection centred on M33. This contamination model also contains the contribution of M31 halo giants to the colour-magnitude diagram (CMD) through isochrone-driven models that encompass the spread in populations through the halo. 

The contamination model is defined over the colour and magnitude ranges $0.2\leq (g_0-i_0)\leq3.0$ and $20 \leq i_0 \leq 24$, and enables the generation of a CMD for contamination at any location in the PAndAS footprint. From these contamination CMDs, we can generate contamination luminosity functions and stellar densities for any region of the PAndAS survey; for full details, see \citet{Martin2013}.

\begin{figure*}
\centering
\includegraphics[width=0.85\textwidth,clip=true]{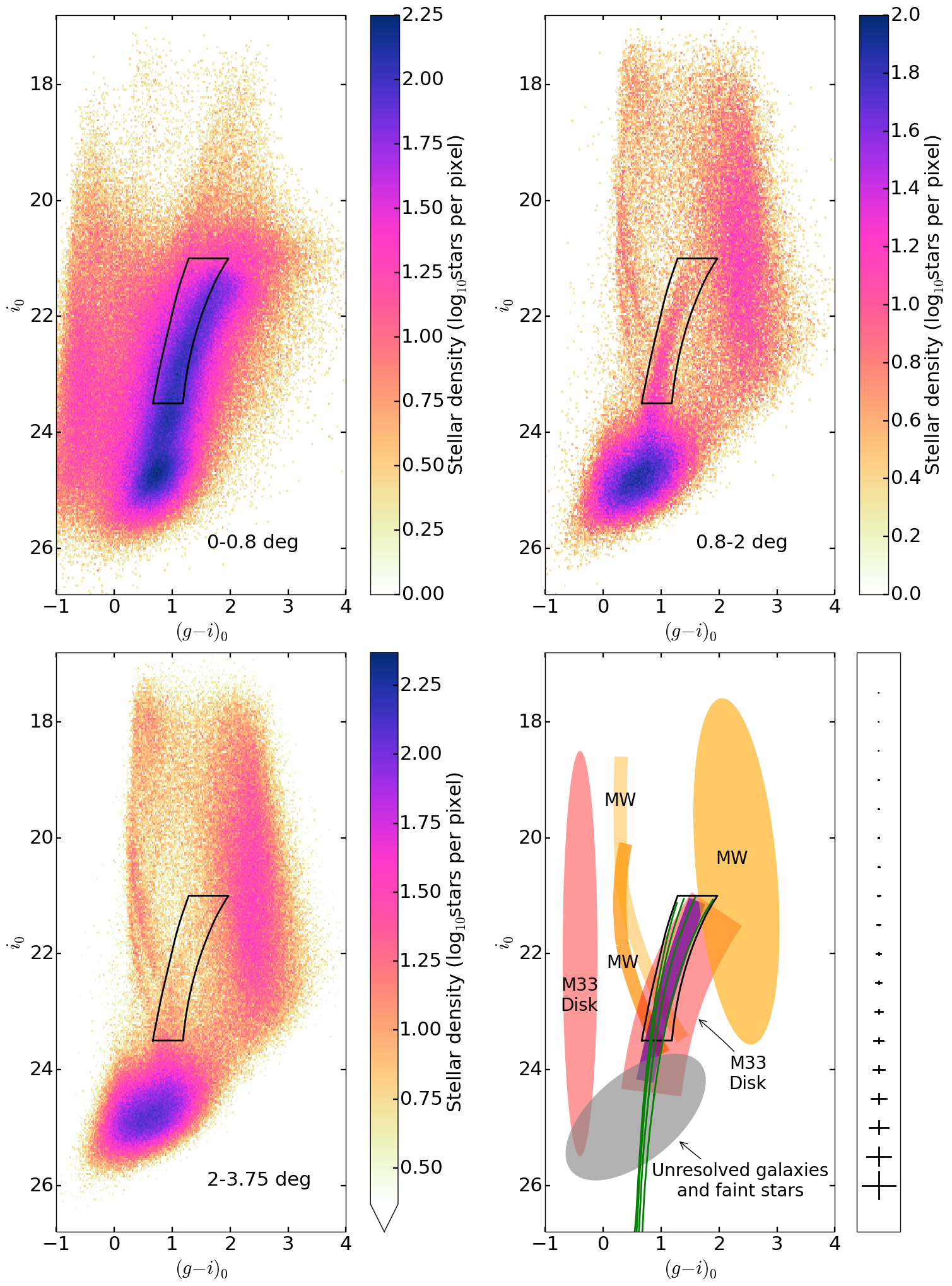}
\caption{Hess diagrams for the three key radial ranges around M33: disk (upper left panel), substructure (upper right panel), and halo (lower left panel). The lower right panel presents a schematic of all key features visible in the other panels. This includes the contamination from the Milky Way (orange), the M33 disk (red), compact background galaxies and faint stars (grey), and the extended stellar substructure around M33 (purple). Isochrones are overlaid in green for ancient $12$~Gyr stellar populations at the distance of M33, with $[\alpha/$Fe$]=0.0$ and with $[$Fe$/$H$]$ ranging from $-2.5$ (leftmost) to $-1.0$ (rightmost) in increments of $0.5$. Mean errors as a function of magnitude for the full stellar population are shown to the right of the lower right panel. The selection box is marked in black on all panels. Pixel size is $0.025 \times 0.033$~mag. Throughout this paper, colour maps were generated using the `cubehelix' scheme \citep{Green2011}.} 
\label{fig:m33_cmd} 
\end{figure*}

\section{Methodology}\label{sec:methods}
In this section, we explain how we have selected the data to be fit, involving colour-magnitude cuts and spatial cuts. Then we detail the process used to fit the data, which involves pixelation of the data, model selection and parametrisation, and Markov Chain Monte Carlo (MCMC) algorithms. Finally, and most importantly, we discuss the synthetic datasets generated to test the reliability of the results of the fitting algorithms.

Figure \ref{fig:m33_cmd} shows Hess diagrams for the region around M33 split into three rough radial ranges: disk, substructure, and halo. The upper left panel covers the inner range (`disk') from the centre out to $0.8$ degrees, within which the disk of M33 dominates completely. The upper right panel covers the intermediate range (`substructure'), from $0.8$ to $2$ degrees. The extended stellar substructure around M33 discussed in \citet{McConnachie2010} is principally contained within this range, and can be seen in the black selection box. The lower left panel contains the outer range (`halo'), from $2$ out to $3.75$ degrees. This range is dominated by contamination, and excluding the extended substructure, is otherwise identical to the `substructure' range. To highlight the similarities between these two ranges, the colour-axis for the `halo' range has been shifted to match the `substructure' range to normalise for area.

We do not extend the outer range past $3.75$ degrees for several reasons; the PAndAS footprint only extends past $3.75$ degrees towards M31 in the North-West, so going beyond this range would bias the data to this quadrant ($\sim25$ per cent of the azimuthal range). For the same reason, this would risk a substantial increase in contamination from the M31 stellar halo, which rapidly increases in density in this direction. Finally, this cut-off matches C13, enabling a more direct comparison to their results.

To assess the stellar completeness within our selection box over the M33 region, we refer the reader to the extensive study presented in Ibata et al. (2014). As noted in this study, the median depth of PAndAS is 26.0 and 24.8 in g and i respectively (5$-\sigma$ detection), while Figure 2 in this paper demonstrate the low dust extinction over the region. Figure 3 in Ibata et al. (2014) presents the spatial completeness (again to 5$-\sigma$) for g (upper) and i (lower) respectively. Noting that the i-band limit and colour range of the selection box presented in Figure \ref{fig:m33_cmd}, we can conclude that the stellar data under examination is complete.

CFHT isochrones were generated using the Dartmouth Stellar Evolution Database\footnote{http://stellar.dartmouth.edu/models/index.html} \citep{Dotter2008} for ancient metal-poor stellar populations at the distance of M33 using a distance modulus of $24.57$ \citep{Conn2012} which corresponds to a distance of $820$~kpc. The selection box marked in black in Figure \ref{fig:m33_cmd} is chosen to cover the only area in colour-magnitude space these isochrones occupy which is not dominated by contamination. The isochrone populations fade to the upper right, as the contamination from the Milky Way rapidly begins to dominate. Beneath the isochrones, we fall into the noise of the background galaxies. The base of the selection box is sufficiently bright so as to avoid issues with spatially variable incompleteness. This selection box roughly matches C13, further enabling a consistent match to their results.

This selection region is unsurprisingly completely overlapped by the M33 disk population. As we are interested in searching for the halo of M33, this dominance at very small radii is not an issue. However, the extended stellar substructure also lies completely on top of this selection box, and is spatially distributed throughout much of the PAndAS footprint around M33, extending beyond $2$~deg. Since this substructure is coincident with the stellar halo both in spatial location and colour-magnitude space, it is impossible to distinguish using photometric data alone. This is very problematic, and complicates the process of detecting the stellar halo significantly.

This selection region is also intersected by two sequences of Milky Way contamination, previously identified in \citet{Martin2014}. The upper sequence was determined to be dominated by the thin Pisces/Triangulum globular cluster stream, independently discovered in the Sloan Digital Sky Survey (SDSS) data by \citet{Bonaca2012} and \citet{Martin2013b}. The lower sequence is likely related to TriAnd2, discovered in \citet{Martin2007}. These streaks are much fainter than the other sources of contamination, however they intersect with the selection box at the base, where the signal for the stellar halo is most likely to be seen.

Any stellar halo will be centrally concentrated, becoming systematically fainter into the outer halo. This fact highlights the difficulty with detecting the M33 halo, as the entire inner region is polluted with the extended substructure, masking the presence of the halo. Looking beyond the substructure in the `halo' range, there is no obvious sign of a halo sequence in the selection box. Therefore, the only hope for finding the halo is a statistical detection. 

\begin{figure}
\centering
\includegraphics[width=0.5\textwidth,clip=true]{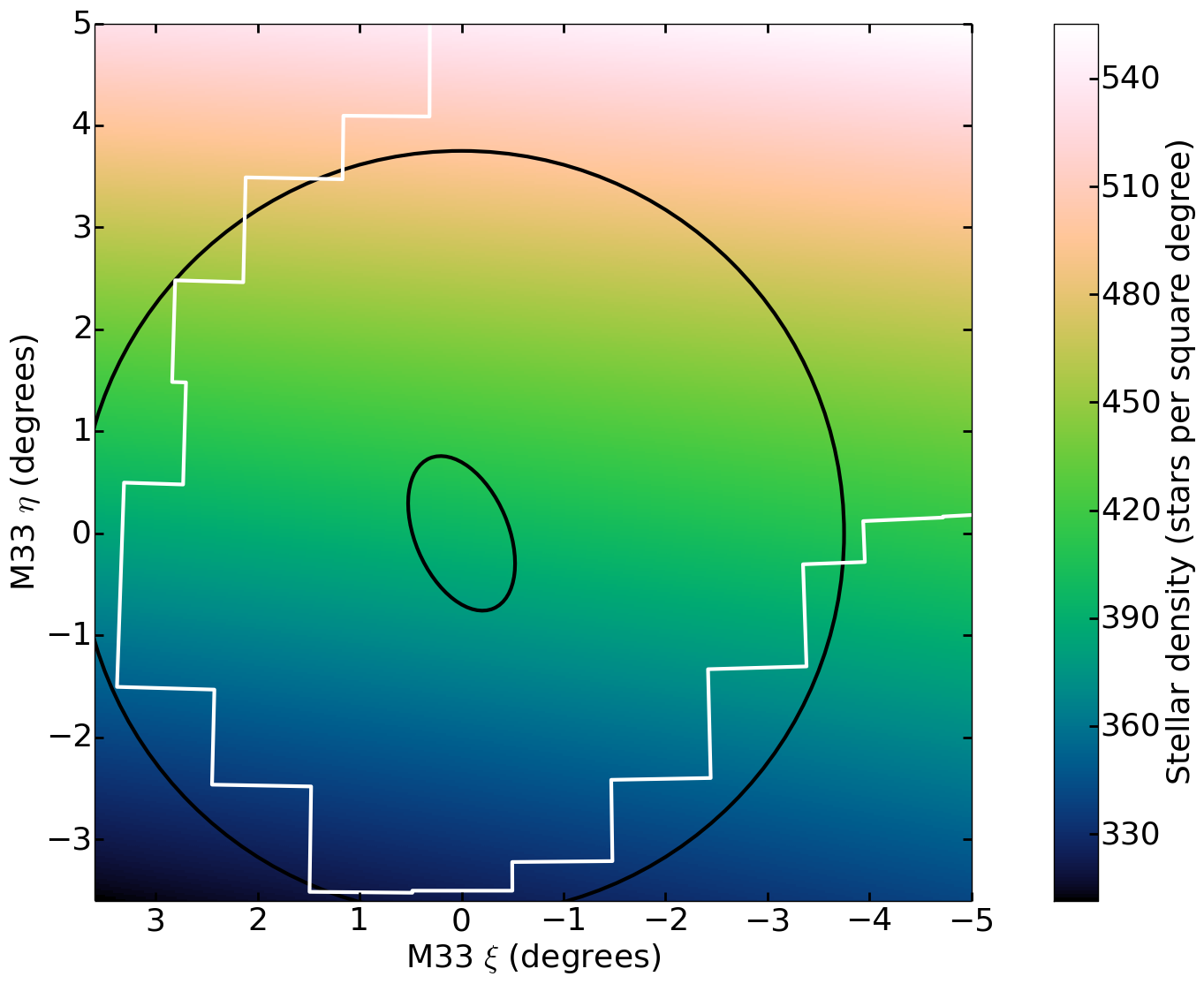}
\caption{Stellar density of the contamination within our colour-magnitude selection box, in the tangent-plane projection centred on M33, generated by the model in Section \ref{sec:contamination}. }
\label{fig:m33_contamination_density} 
\end{figure}

By summing the contamination model over the colour-magnitude selection box, we obtain a stellar density contamination map for the entire M33 region, shown in Figure \ref{fig:m33_contamination_density}. This contamination map varies by $50$ per cent over our spatial selection, so it is a significant shift away from the assumption of a constant contamination level used in C13.

\begin{figure*}
\centering
\includegraphics[width=0.9\textwidth,clip=true]{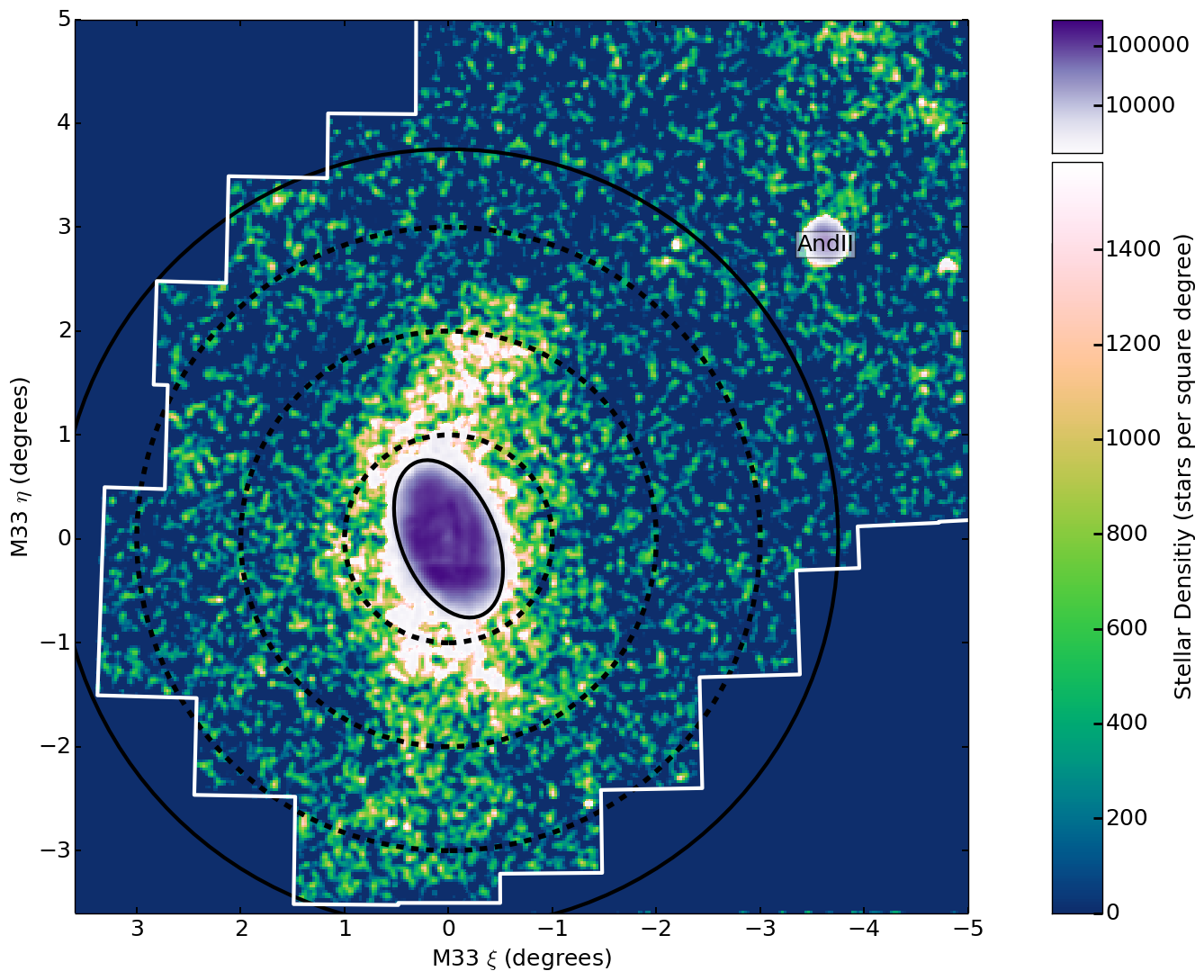}
\caption{Stellar density map of PAndAS stars with de-reddened colours and magnitudes consistent with metal-poor red giant branch populations at the distance of M33, using the selection box in Figure \ref{fig:m33_cmd}. Pixels are $0.025\degr \times 0.025\degr$, and the map has been smoothed (to improve visual presentation) using a Gaussian with a dispersion of $\sigma = 0.1\degr$. It is displayed in tangent-plane projection centred on M33, with scaling chosen to highlight the structure in the halo region. This map has been contamination-subtracted, using the model discussed in Section \ref{sec:contamination} (see \citealt{Martin2013} for full details). The dashed lines mark the $1$, $2$, and $3$~deg radii, and the solid black circle marks the $3.75$~deg radius. The M33 disk is marked by a black ellipse, and the boundary of the PAndAS footprint is marked by a solid white line.}
\label{fig:m33_smooth_map} 
\end{figure*} 

A smoothed contamination-corrected stellar density map for the M33 region, using the colour-magnitude selection box discussed above, is shown in Figure \ref{fig:m33_smooth_map}. The M33 disk dominates in the centre, with the extended substructure dominating the rest of the innermost radial degree, and stretching out beyond two degrees to the North, and out to two degrees to the South. M31 lies $\sim15$~deg to the North-West, with the prominent dwarf galaxy Andromeda II nearby, but outside our selection radius, at ($-3.6,2.8$). Andromeda XXII is within our selection but barely visible at ($-1.4,-2.6$). Several globular cluster systems of background galaxies are visible within our selection around ($-2.5,3.0$) \citep{Martin2013}. Ideally, any substructures such as these will be returned by the fitting algorithm.

The structure in the M33 disk seen in the central parts of Figure \ref{fig:m33_smooth_map};  while the spiral structure of the disk are apparent, other features (white blurred lines) are apparent due to overcrowding in the inner regions of the disk.
Despite our resolve to avoid masking in general, the significant crowding problems within the disk, along with other blemishes, force us to mask out the entire disk region to an elliptical radius of $0.8$~deg, with an ellipticity of $0.5$ and position angle of $23$~deg. 
The ellipticity, $e$, is defined $e = 1 - b/a$, where $b$ and $a$ are the semi-minor and semi-major axes of the ellipse, respectively. The position angle
is taken from north towards east (counter-clockwise, in the figures presented throughout this work).

\subsection{Data Pixelation}\label{sec:pixelation}
We now pixelate all the stars within the colour-magnitude selection box (ignoring any spatial restrictions). This enables the calculation of residuals from models, which will be a key analysis tool later in this paper. We note here that the focus is on replicating C13. Additional algorithms were developed, but these didn't have any further success detecting the halo; we include them in the Appendix for completeness. 

\begin{figure}
\centering
\includegraphics[width=0.5\textwidth,clip=true]{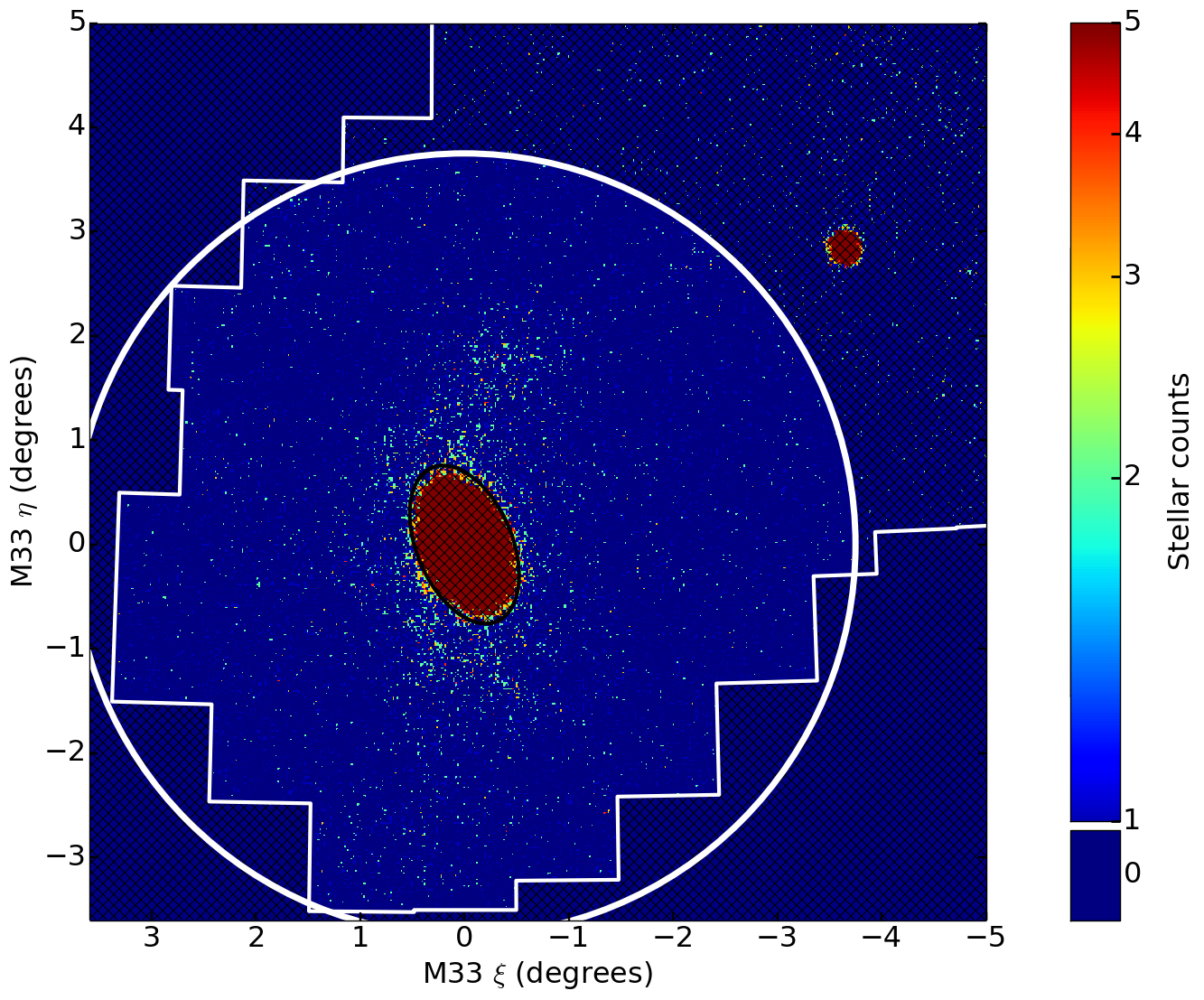}
\caption{Ultra-fine pixelation of the stellar data within the colour-magnitude selection box using a logarithmic colour-axis. The solid white circle marks the $3.75$~deg radius. The M33 disk mask is marked by a black ellipse, and the boundary of the PAndAS footprint is marked by a solid white line. Pixels excluded from the fits in the rest of this work are marked with black cross-hatching. Pixels are $0.02\degr \times 0.02\degr$.}
\label{fig:m33_430} 
\end{figure}
\begin{figure}
\centering
\includegraphics[width=0.5\textwidth,clip=true]{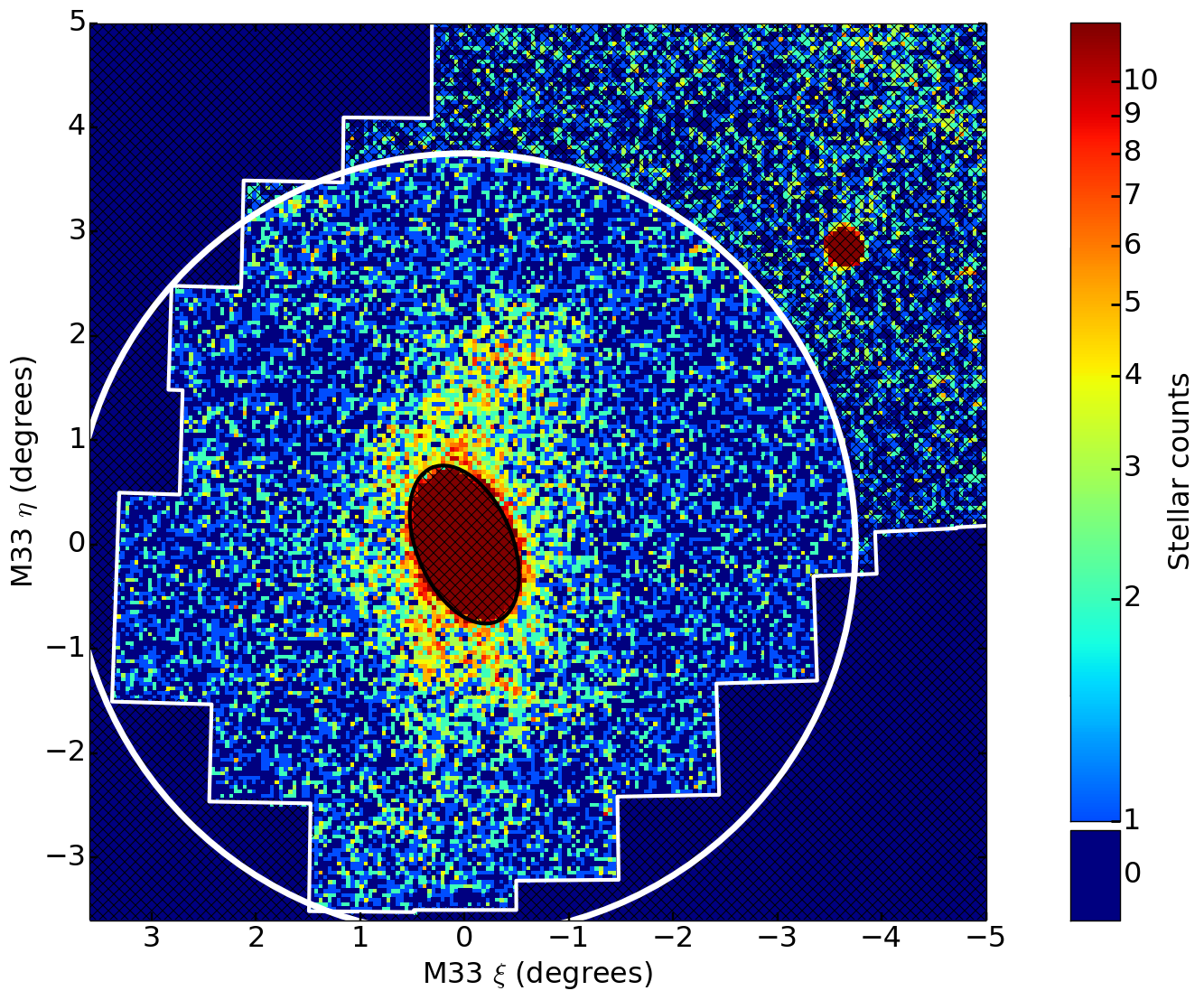}
\caption{Fine pixelation of the stellar data within the colour-magnitude selection box using a logarithmic colour-axis. The solid white circle marks the $3.75$~deg radius. The M33 disk mask is marked by a black ellipse, and the boundary of the PAndAS footprint is marked by a solid white line. Pixels excluded from the fits in the rest of this work are marked with black cross-hatching. Pixels are $0.043\degr \times 0.043\degr$.}
\label{fig:m33_200} 
\end{figure}
\begin{figure}
\centering
\includegraphics[width=0.5\textwidth,clip=true]{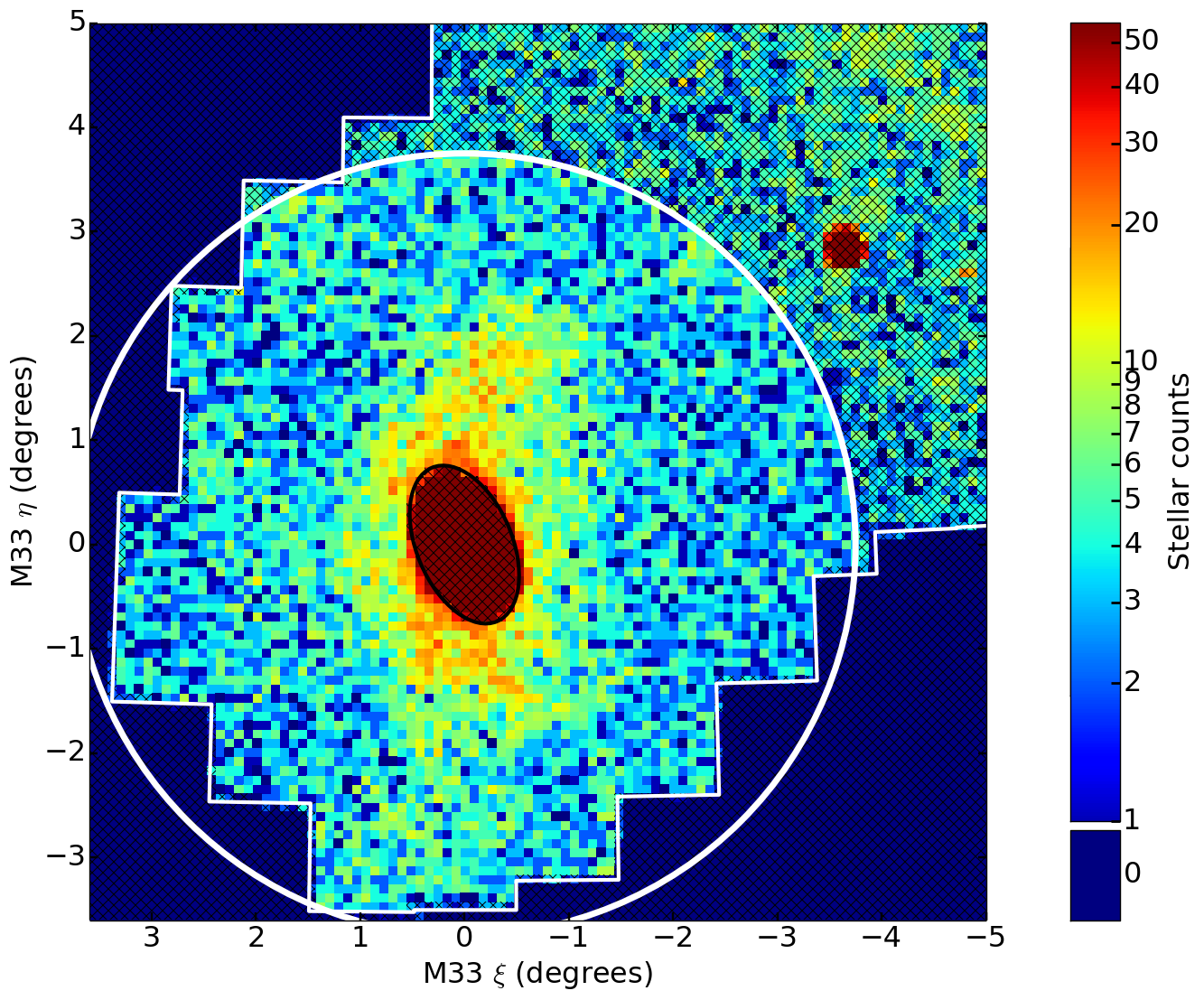}
\caption{Coarse pixelation of the stellar data within the colour-magnitude selection box using a logarithmic colour-axis. The solid white circle marks the $3.75$~deg radius. The M33 disk mask is marked by a black ellipse, and the boundary of the PAndAS footprint is marked by a solid white line. Pixels excluded from the fits in the rest of this work are marked with black cross-hatching. Pixels are $0.087\degr \times 0.087\degr$.}
\label{fig:m33_100} 
\end{figure}

The three pixelations used for the rest of this work are shown in Figures \ref{fig:m33_430}, \ref{fig:m33_200}, and \ref{fig:m33_100}. Following C13, the ultra-fine pixelation $430 \times 430$ with pixel sizes of $0.02\degr \times 0.02\degr$. In addition to this we generate two more pixelations; the fine pixelation is $200 \times 200$, with pixel sizes of $0.043\degr \times 0.043\degr$, and the coarse pixelation is $100 \times 100$, with pixel sizes of $0.087\degr \times 0.087\degr$. In all cases, any pixels even partially outside the PAndAS footprint are excluded from fits, as well as any pixels which have centres within the M33 disk mask or outside the $3.75$~deg radius. 

Each of these pixelations is useful in different ways. The coarse pixelation clumps substructure together, allowing the fitting algorithms to more easily detect the presence of substructure in any single pixel. In contrast, the fine pixelation provides a good spatial resolution for all the substructure, however the increased pixel density also comes at the cost of a longer run-time for the algorithms. Furthermore, these two pixelations provide a consistency check for the algorithms, as they should produce the same results. The ultra-fine pixelation enables a direct comparison to C13; at this resolution most pixels are empty, and the clumpy nature of the substructure can be seen.

Pixelations coarser than our $100 \times 100$ grid discard too much spatial information. Conversely, pixelations finer than our $200 \times 200$ grid are also problematic, not just due to time and space complexity issues for the algorithms, but because they provide no smoothing on the scale of gaps in the dataset. While these gaps are in theory covered by fake entries in the catalogue, in practice this is an imperfect process. Residual comparisons also lose their visual utility, as pixel counts fall to zero in most pixels. For the fine pixelation in Figure \ref{fig:m33_200}, the majority of pixels usable for fits are already $\leq1$ star per pixel. These problems are unfortunately unavoidable for the ultra-fine pixelation.

\subsection{Models}\label{sec:models}
Modelling the stellar halo as a projected generalised triaxial ellipsoid would require many parameters and would greatly complicate the fitting procedure. Instead, we simply assume the stellar halo is spherically symmetric and centred on M33, enabling us to model it with a simple two parameter exponential radial profile. Thus, combined with a contamination component, the data is modelled by
\begin{equation}\label{eq:models}
\Sigma_{(\xi,\eta)} = \Sigma_0\exp\left(-\frac{r_{(\xi,\eta)}}{r_s}\right) + \Sigma_{C(\xi,\eta)},
\end{equation}
where $\Sigma_0$ varies the intensity of the halo component, $r_s$ is the scale length of the halo, and $\Sigma_{C(\xi,\eta)}$ is the contamination model as discussed in Section \ref{sec:contamination}. We note that fits to stellar halos typically consider power-law distributions (e.g. \citealt{2004MNRAS.347..556Z,2005astro.ph..2366G,2008ApJ...685L.121B,2014ApJ...787...30D}), but here an exponential is chosen to allow a robust comparison with C13 who also employed this model. Furthermore, given the expected low signal-to-noise of the signal of any stellar halo (see Figure 7 in C13), the differences in an exponential and power-law distribution between one and three scale-lengths will be effectively indistinguishable.

Several other models for the contamination were also considered, including a simple constant $\Sigma_C$ (as was used in C13), a combination of a constant and the spatially varying contamination model $\Sigma_{C(\xi,\eta)} + \Sigma_C$, and a modified version of the spatially varying contamination model with an extra free parameter allowing the intensity to vary $\alpha\Sigma_{C(\xi,\eta)}$; this last model was to allow for the possibility that the spatially varying contamination model was not suitability optimised for the M33 region. All of these produce equivalent or inferior fits to the original spatially varying contamination model $\Sigma_{C(\xi,\eta)}$, thus for the remainder of this work, all fits use this contamination model. 

\subsection{MCMC Algorithms}
The backbone of all of the fitting algorithms we use is a pseudo-randomised walk through the parameter space via a MCMC, an algorithm that efficiently samples the parameters space to give a measure of the posterior distribution. A simple example of a MCMC is the Metropolis-Hastings algorithm, which pseudo-randomly steps though parameter space, evaluating the likelihood function on each step to determine the chance of accepting that step.

Simple algorithms such as this suffer greatly when the parameter space contains strong degeneracies, which relegate most of the parameter space and thus most of the attempted steps within it, to very low probability. 
Even with the simple parametrisation of the model we have chosen, there is a strong degeneracy between the halo parameters $\Sigma_0$ and $r_s$. 

\citet{Goodman2010} proposed an affine-invariant ensemble sampler for MCMC, which is able to sample parameter spaces with strong degeneracies with much greater efficiency $-$ over an order of magnitude faster than Metropolis-Hastings algorithms. The key to its method is instead of sending out a single walker, it sends out a large ensemble of walkers, which sample information from each other to determine which points to investigate in parameter space.

We use this affine-invariant sampler for all of our algorithms, with the exception of the final algorithm discussed in this work. Due to the complexity of this last method, we fall back on a standard Metropolis-Hastings algorithm in this case.

\subsection{Synthetic Data}

\begin{figure}
\centering
\includegraphics[width=0.5\textwidth,clip=true]{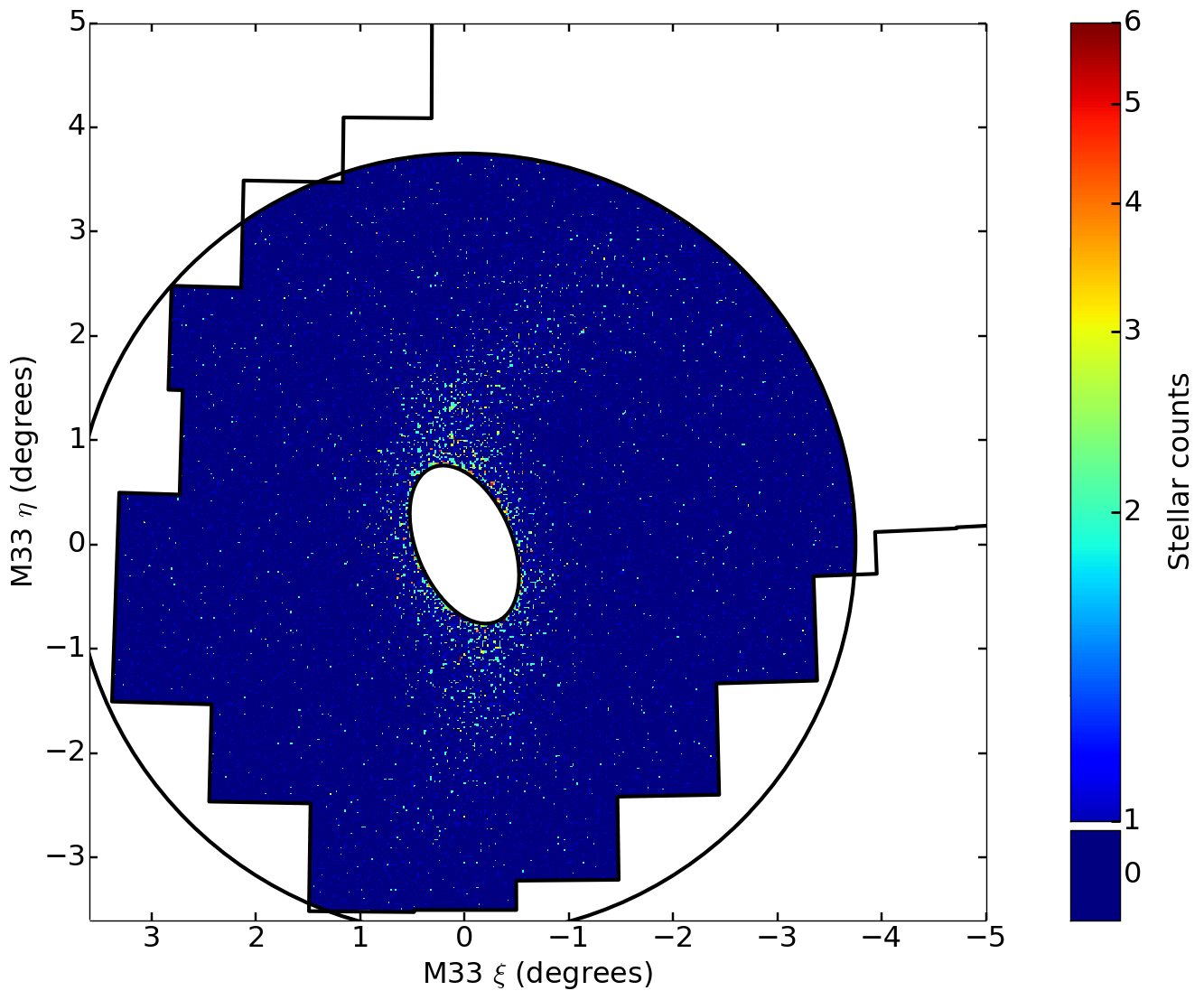}
\caption{Random realisation of the synthetic data, including extended substructure, designed to match the ultra-fine pixelation of the stellar data in Figure \ref{fig:m33_430}, using a logarithmic colour-axis. The solid black circle marks the $3.75$~deg radius. The M33 disk mask is marked by a black ellipse, and the boundary of the PAndAS footprint is marked by a solid black line. Pixels are $0.02\degr \times 0.02\degr$.}
\label{fig:syn_430} 
\end{figure}
\begin{figure}
\centering
\includegraphics[width=0.5\textwidth,clip=true]{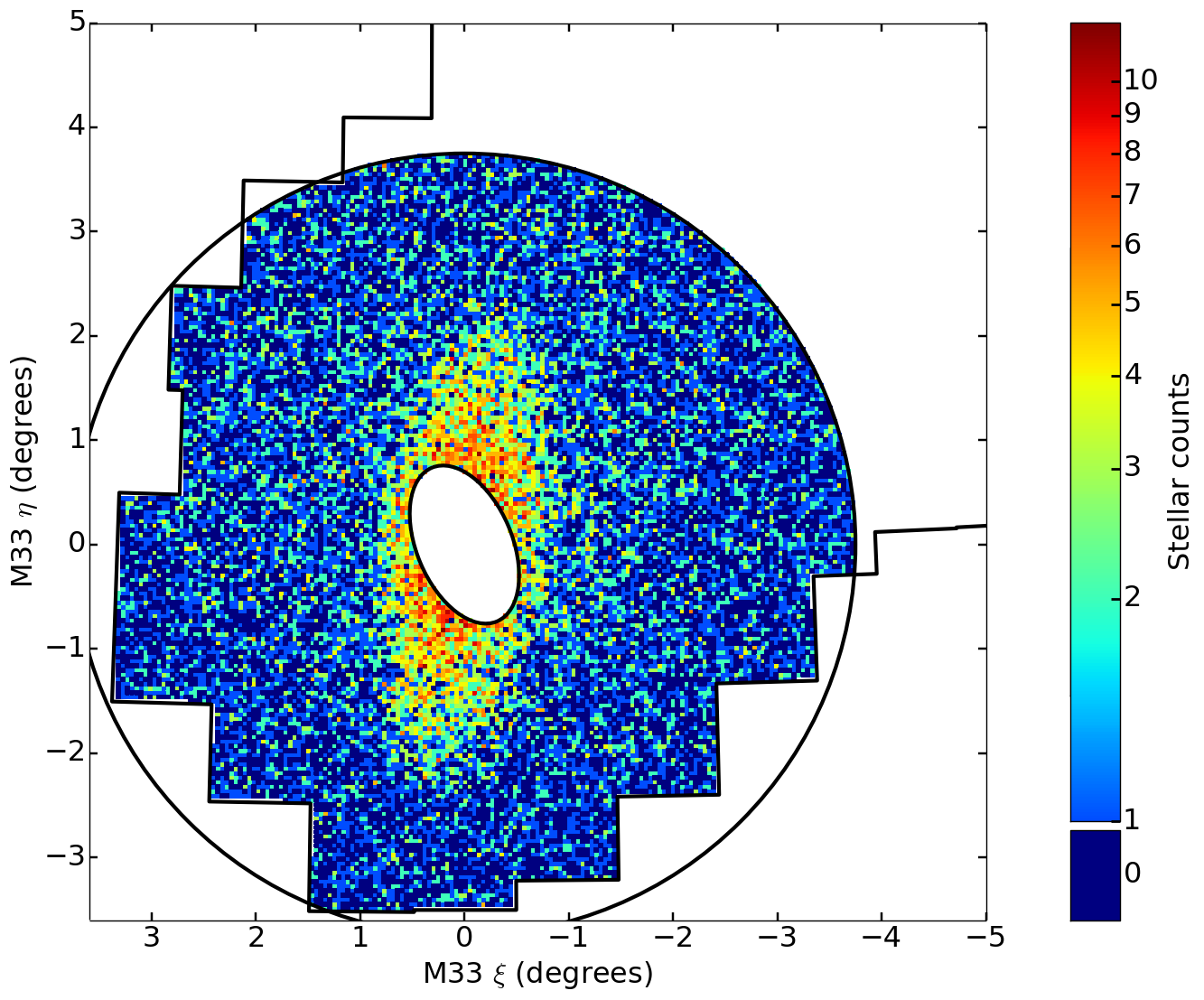}
\caption{Random realisation of the synthetic data, including central substructure, designed to match the fine pixelation of the stellar data in Figure \ref{fig:m33_200}, using a logarithmic colour-axis. The solid black circle marks the $3.75$~deg radius. The M33 disk mask is marked by a black ellipse, and the boundary of the PAndAS footprint is marked by a solid black line. Pixels are $0.043\degr \times 0.043\degr$.}
\label{fig:syn_200} 
\end{figure}
\begin{figure}
\centering
\includegraphics[width=0.5\textwidth,clip=true]{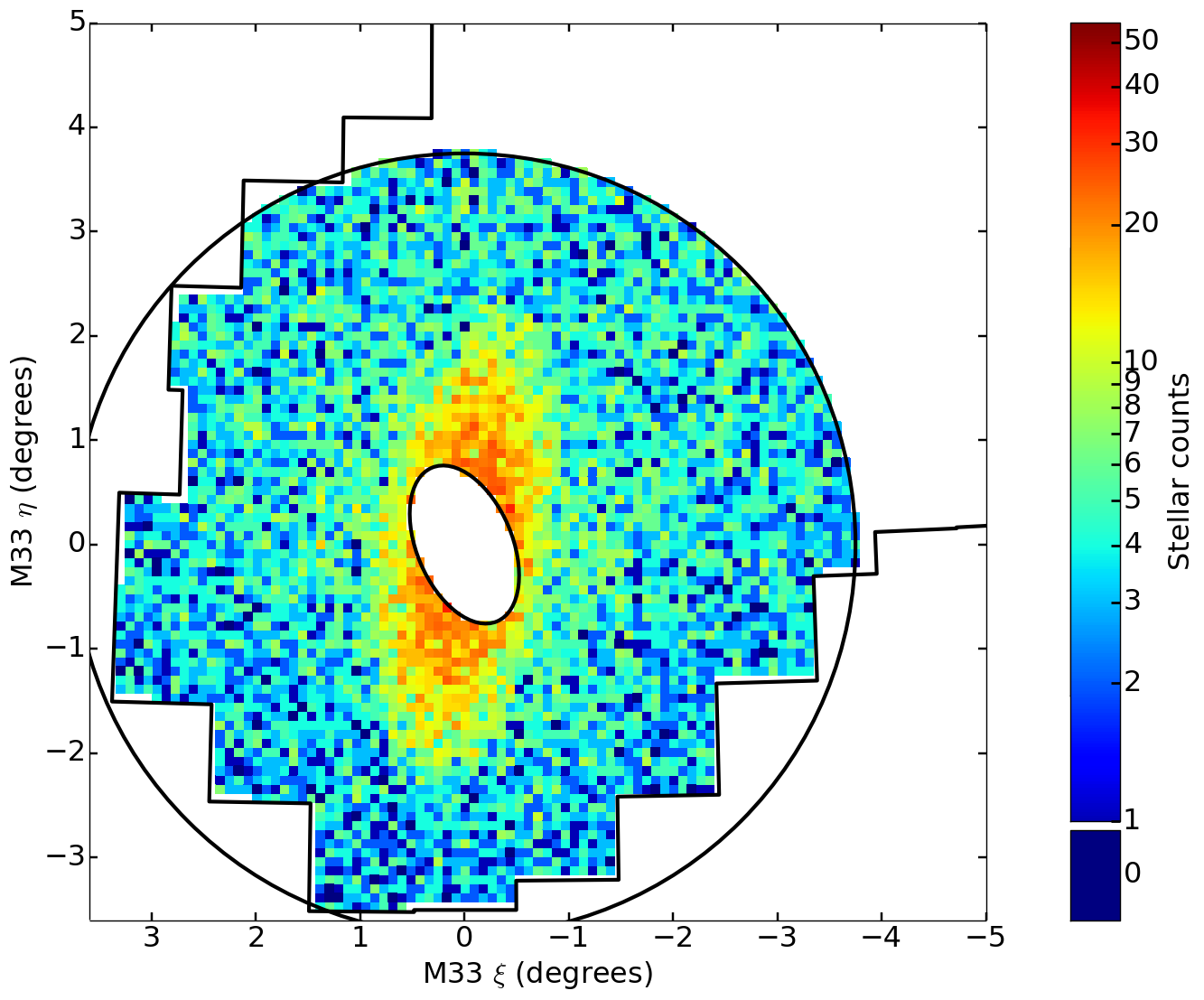}
\caption{Random realisation of the synthetic data, including central substructure, designed to match the coarse pixelation of the stellar data in Figure \ref{fig:m33_100}, using a logarithmic colour-axis. The solid black circle marks the $3.75$~deg radius. The M33 disk mask is marked by a black ellipse, and the boundary of the PAndAS footprint is marked by a solid black line. Pixels are $0.087\degr \times 0.087\degr$.}
\label{fig:syn_100} 
\end{figure}

The most important aspect of this analysis, is our use of synthetic data testing, or `Sanity Testing'. We generate mock datasets with similar structures to the PAndAS data for M33, and then test the precision and accuracy of the fitting algorithms in retrieving the known parameters. This step provides critical verification as to how strongly (or not) the results from a given fitting method can be trusted. 

To optimally test the validity of the algorithms for the pixelations in Section \ref{sec:pixelation}, the synthetic data must be a close match. Thus, we use the same grids, boundaries, masks and selections of usable pixels as for the PAndAS data. Using the same centre of M33, and the model in Equation \ref{eq:models} combined with a model for the substructure, we generate noise-free synthetic pixelations. Finally, to add noise we use the value in each pixel, $n_{(\xi,\eta)}$, to define a Poisson distribution with this as the mean value, 
\begin{equation}
P(N) = \frac{n_{(\xi,\eta)}^{N} \exp(-n_{(\xi,\eta)})}{N!}\ , 
\end{equation}
and replace the pixel value with a random deviate from this distribution.

To test the validity of any results found through replicating the methodology of C13, we generate a synthetic model to match the ultra-high resolution pixelation in Figure \ref{fig:m33_430}. This model does not contain any halo component; in this way, we test to see if a halo can be recovered by the algorithm when none is present. To accommodate a good match to the data in the absence of a halo, we use an elongated model of the substructure. This structure is based on a sum of Gaussian components, so bleeds into its surroundings, as would be expected for the true substructure.

Alternative algorithms discussed in this work are tested against synthetic data generated to match the coarser pixelations. Unlike the synthetic ultra-fine pixelation, these synthetic pixelations include halo components. 
The best fit halo parameter values reported in C13 are $\Sigma_0 = 158 \pm 83$ stars per square degree for the halo central density and $r_s = 1.5 \pm 1.3$~deg for the scale radius. For the synthetic datasets we use a slightly smaller scale radius of $1.2$~deg, for a better visual match to the pixelated data, which is still well within the uncertainty range given in C13. For the central density, we use a much brighter value of $800$ stars per square degree. This provides a reasonable match to the data, and at approximately $5$ times brighter than the value reported in C13, ensures that any method capable of detecting the stellar halo from C13 will also detect the halo in the synthetic data. Thus any methods which are unable to correctly recover the halo parameters in the synthetic pixelations, can be safely excluded as viable candidates for use with the PAndAS data.

It is important to include the central substructure in all the synthetic data, as it dominates at small radii, where the signal of the halo is expected to be strongest. All other substructure is excluded for simplicity, due to ignorance of the true substructure, and to maximise the chance of the fitting method to detect the halo in the synthetic data. For the coarse and fine pixelations, we simplify the extended substructure, representing it very roughly as an ellipse centred on M33 with a position angle of $-8$~deg, ellipticity $0.65$. The density falls linearly by elliptical radius from $3000$ stars per square degree at the centre to zero at $2.5$~deg. We use a linear decay to avoid the contamination from the substructure bleeding into too many of the usable pixels, and also to avoid the form of the model matching the form of the halo model $-$ reducing the chance that the fitter would select the much more dominant signal of the substructure as the halo. These simplifications are made to further aid the algorithms in recovering the halo parameters, thus giving us extra confidence in their inability to successfully recover halo parameters for the PAndAS data in the case of their failure with the synthetic data.

\begin{figure}
\centering
\includegraphics[width=0.5\textwidth,clip=true]{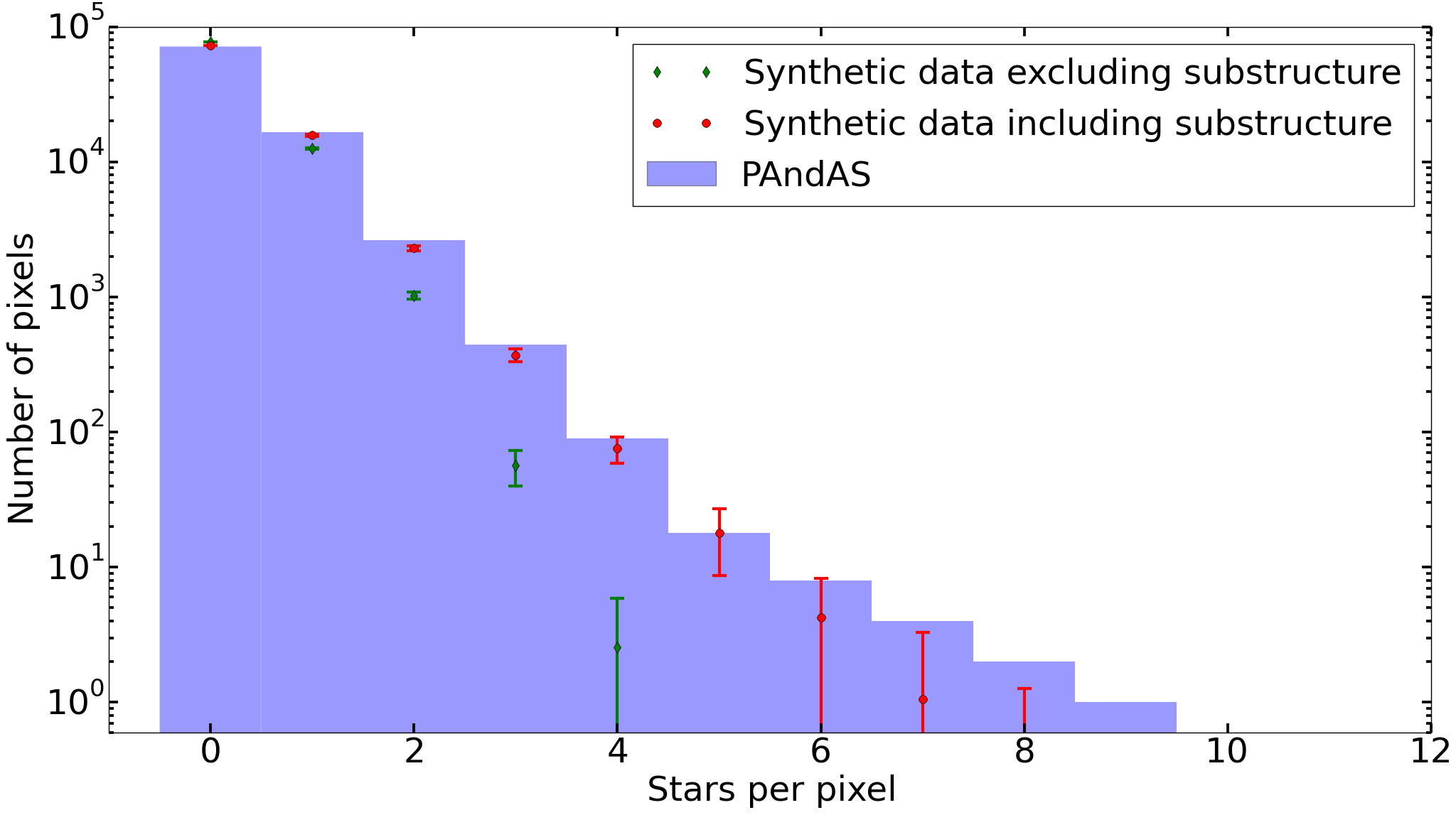}
\caption{Histogram of the pixel values in the ultra-fine pixelation in Figure \ref{fig:m33_430}, compared to $100$ random realisations of the synthetic data as exemplified in Figure \ref{fig:syn_430}. Error bars give the $2 \sigma$ range for the ensemble of realisations of the synthetic data.}
\label{fig:syn_hist_430} 
\end{figure}
\begin{figure}
\centering
\includegraphics[width=0.5\textwidth,clip=true]{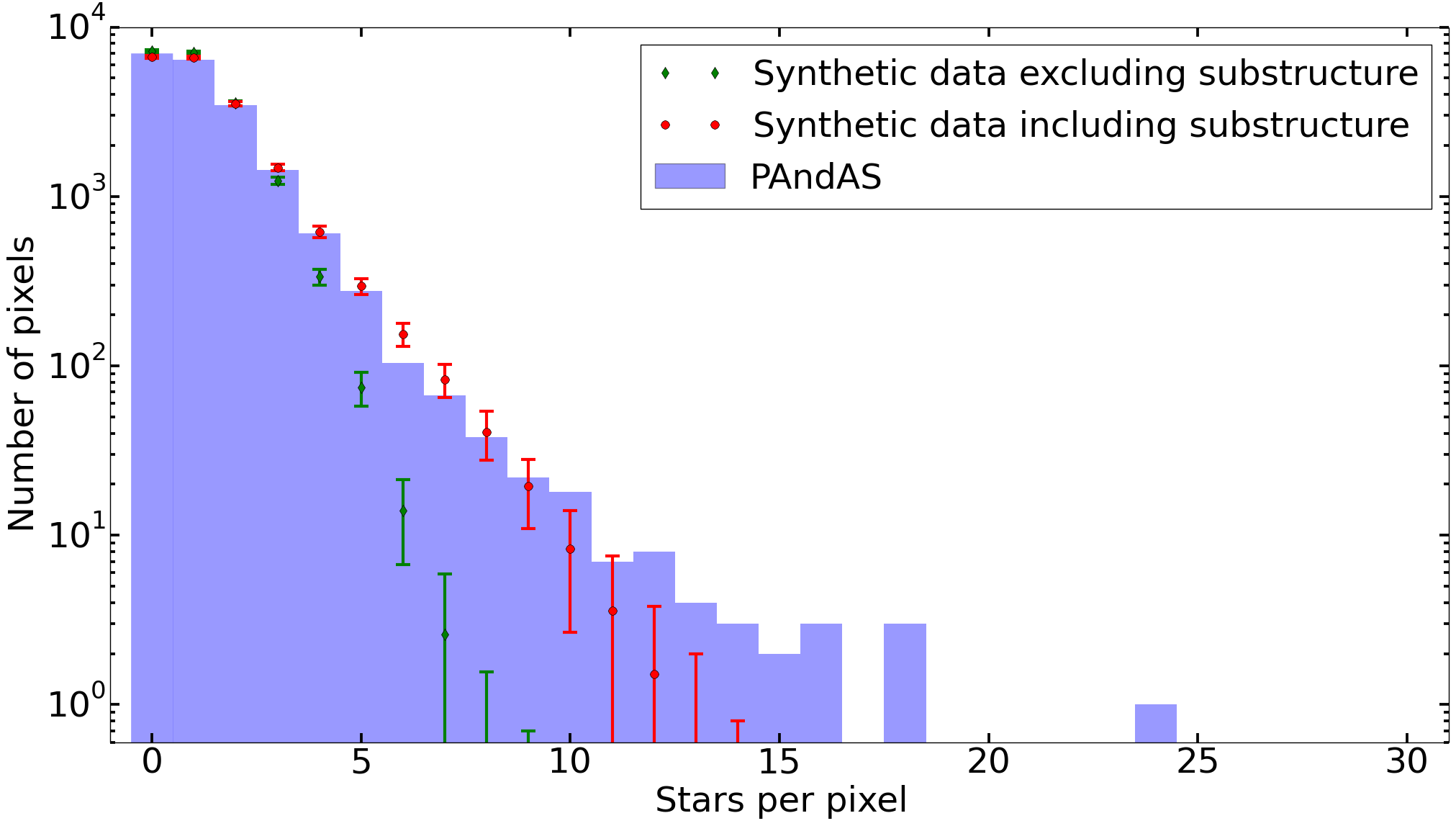}
\caption{Histogram of the pixel values in the fine pixelation in Figure \ref{fig:m33_200}, compared to $100$ random realisations of the synthetic data as exemplified in Figure \ref{fig:syn_200}. Error bars give the $2 \sigma$ range for the ensemble of realisations of the synthetic data.}
\label{fig:syn_hist_200} 
\end{figure}
\begin{figure}
\centering
\includegraphics[width=0.5\textwidth,clip=true]{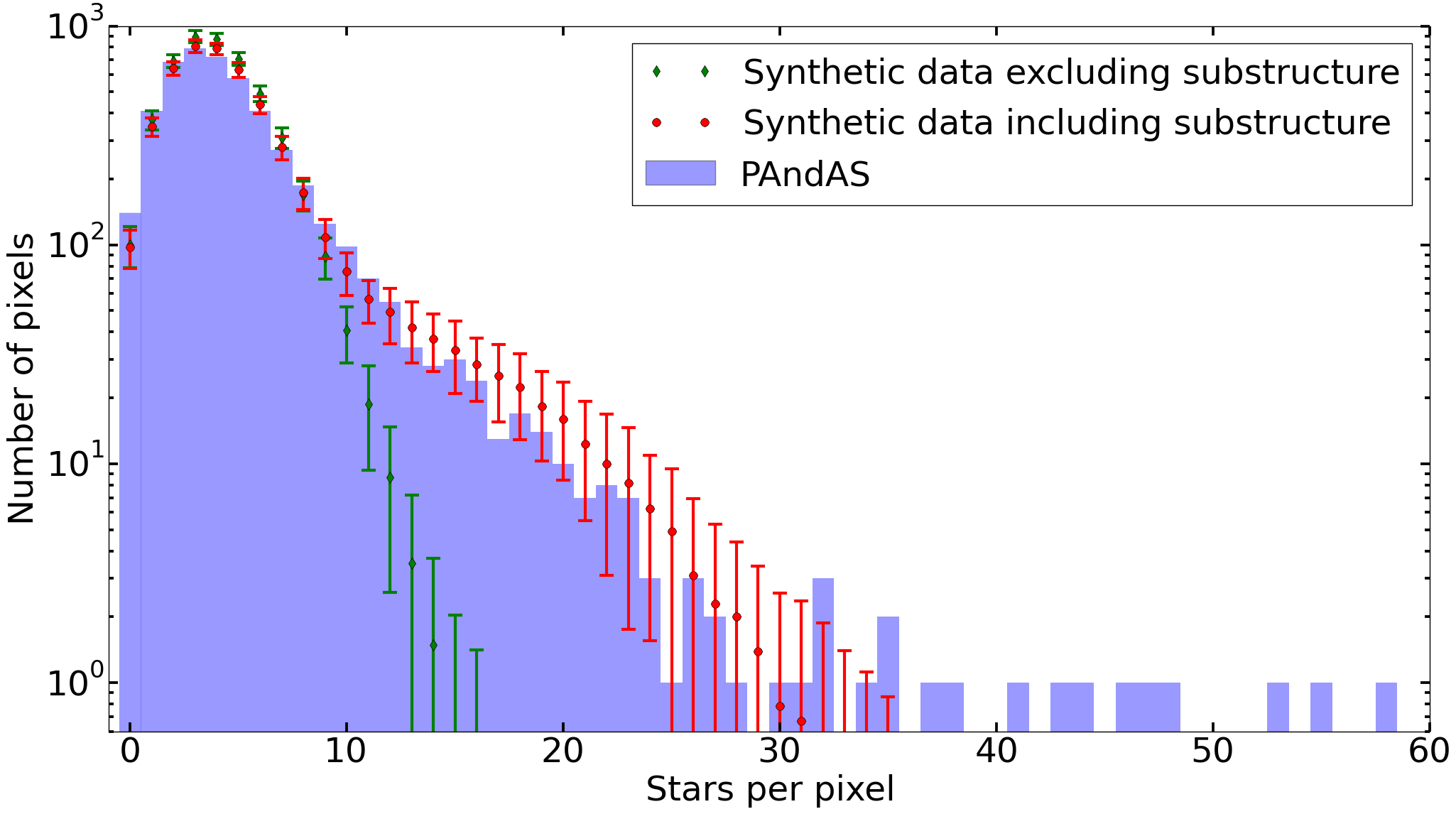}
\caption{Histogram of the pixel values in the coarse pixelation in Figure \ref{fig:m33_100}, compared to $100$ random realisations of the synthetic data as exemplified in Figure \ref{fig:syn_100}. Error bars give the $2 \sigma$ range for the ensemble of realisations of the synthetic data. The horizontal range excludes three pixels with values between $60$ and $100$ stars per pixel.}
\label{fig:syn_hist_100} 
\end{figure}

Figures \ref{fig:syn_430}, \ref{fig:syn_200}, and \ref{fig:syn_100} show random realisations generated using the above method and parameters to match the pixelations in Figures \ref{fig:m33_430}, \ref{fig:m33_200}, and \ref{fig:m33_100}. We also generate substructure-free synthetic datasets for more extreme `Sanity Testing' in the same way. 
We emphasise that the goal of the synthetic data is not to precisely match the true data, but simply to roughly match the data so that the parameter retrieval precision and accuracy of the fitting methods can be tested.
Figures \ref{fig:syn_hist_430}, \ref{fig:syn_hist_200}, and \ref{fig:syn_hist_100} compare histograms of the true data pixelations to $100$ random realisations of the synthetic data for ultra-fine, fine, and coarse pixelations respectively. These histograms demonstrate a reasonable match, with the a slight underestimate of the concentration of the substructure $-$ further ensuring that the halo is detectable in the synthetic data, if at all. 

\section{Revisiting C13 and Beyond}\label{sec:C13}
The  motivation of this work is to characterise the the smooth stellar halo component of M33. Here we replicate the method used by C13, with our updated data for M33 and more detailed contamination. We compare the results to an analysis of similar synthetic data, that is known not to contain a halo component.

\begin{figure}
\centering
\includegraphics[width=0.5\textwidth,clip=true]{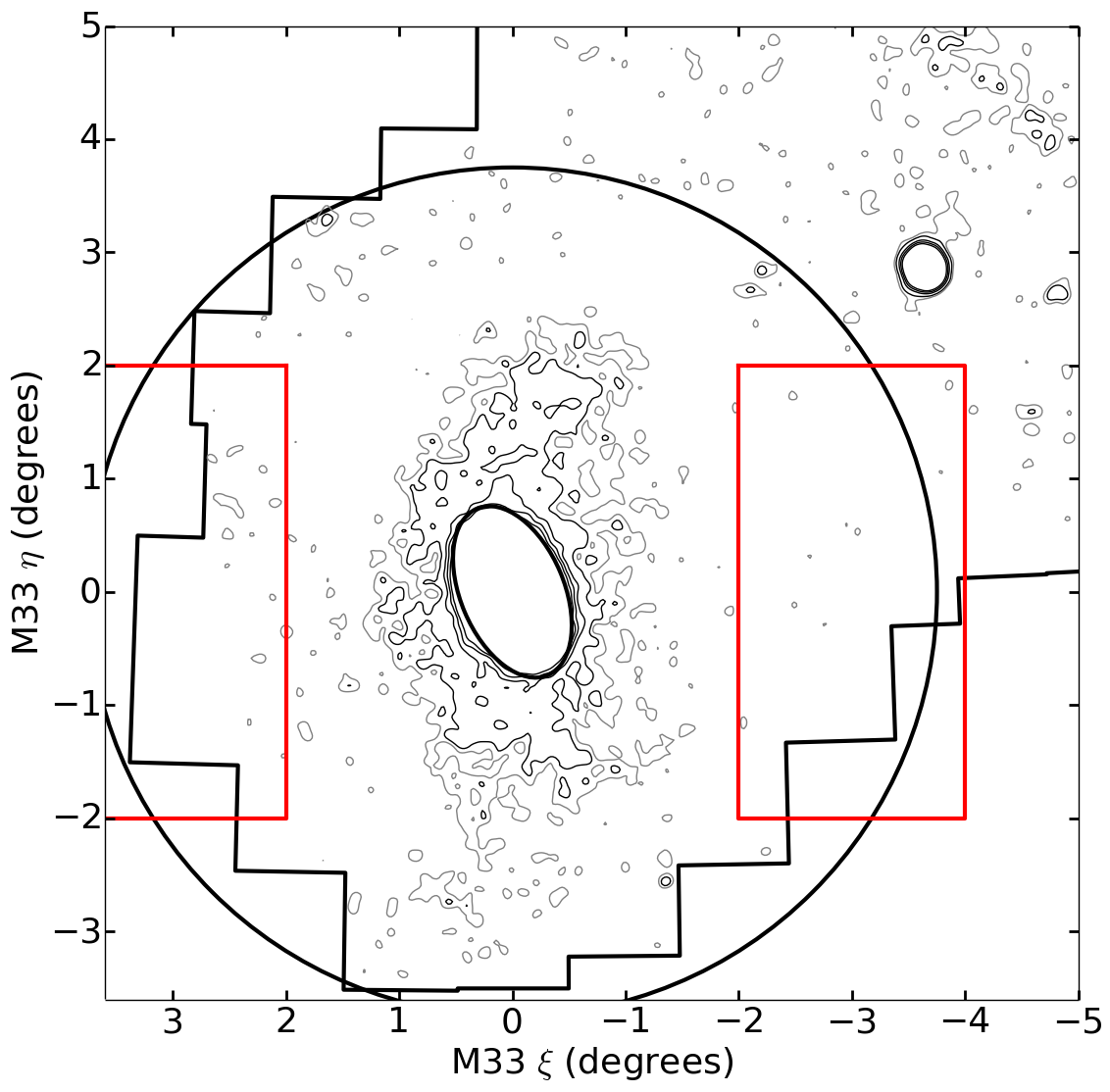}
\caption{Density distribution of PAndAS stars using the ultra-fine pixelation in Figure \ref{fig:m33_430}. The grey contour is 1$\sigma$ above the background. The black contours are 2$\sigma$, 5$\sigma$, 8$\sigma$, and 12$\sigma$ above the background. The regions used for background estimation are marked with red rectangles. The solid black circle marks the $3.75$~deg radius. The M33 disk mask is marked by a black ellipse, and the boundary of the PAndAS footprint is marked by a thick solid black line. Pixels are $0.02\degr \times 0.02\degr$.}
\label{fig:contour_m33} 
\end{figure}

\begin{figure}
\centering
\includegraphics[width=0.5\textwidth,clip=true]{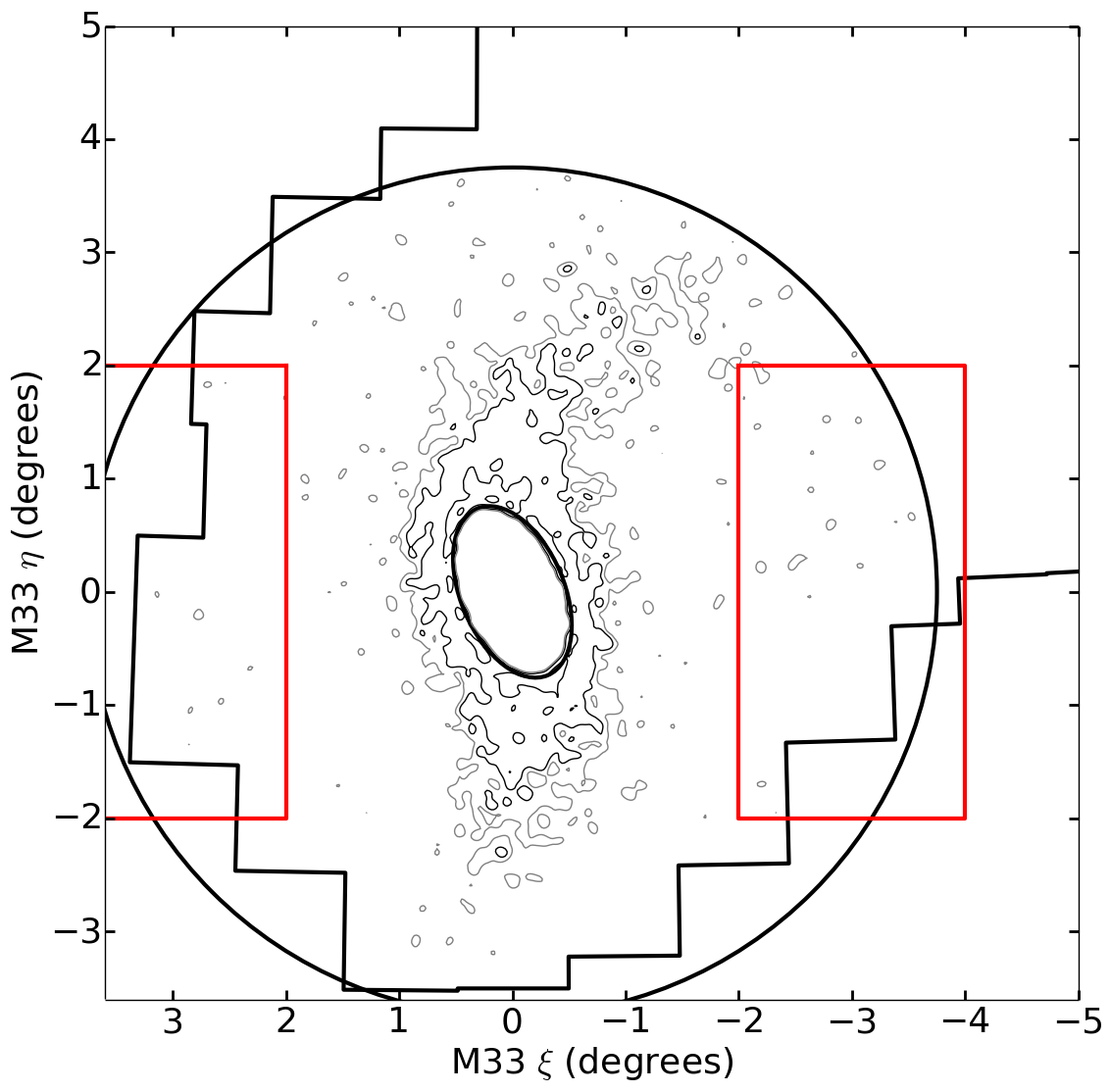}
\caption{Density distribution of synthetic stars using the ultra-fine pixelation in Figure \ref{fig:syn_430}. The grey contour is 1$\sigma$ above the background. The black contours are 2$\sigma$, 5$\sigma$, 8$\sigma$, and 12$\sigma$ above the background. The regions used for background estimation are marked with red rectangles. The solid black circle marks the $3.75$~deg radius. The M33 disk mask is marked by a black ellipse, and the boundary of the PAndAS footprint is marked by a thick solid black line. Pixels are $0.02\degr \times 0.02\degr$.}
\label{fig:contour_syn} 
\end{figure}

Smoothing Figures \ref{fig:m33_430} and \ref{fig:syn_430} using a Gaussian with a dispersion of $\sigma = 0.08\degr$, or four pixel widths, results in Figures \ref{fig:contour_m33} and \ref{fig:contour_syn} respectively. The background level is estimated as the mean of the pixels within the two red rectangles, ignoring any pixels beyond the 3.75 degree cutoff or outside the PAndAS footprint. Contours are drawn to match C13, and show a reasonable match between the PAndAS data and the synthetic data.

\begin{figure}
\centering
\includegraphics[width=0.5\textwidth,clip=true]{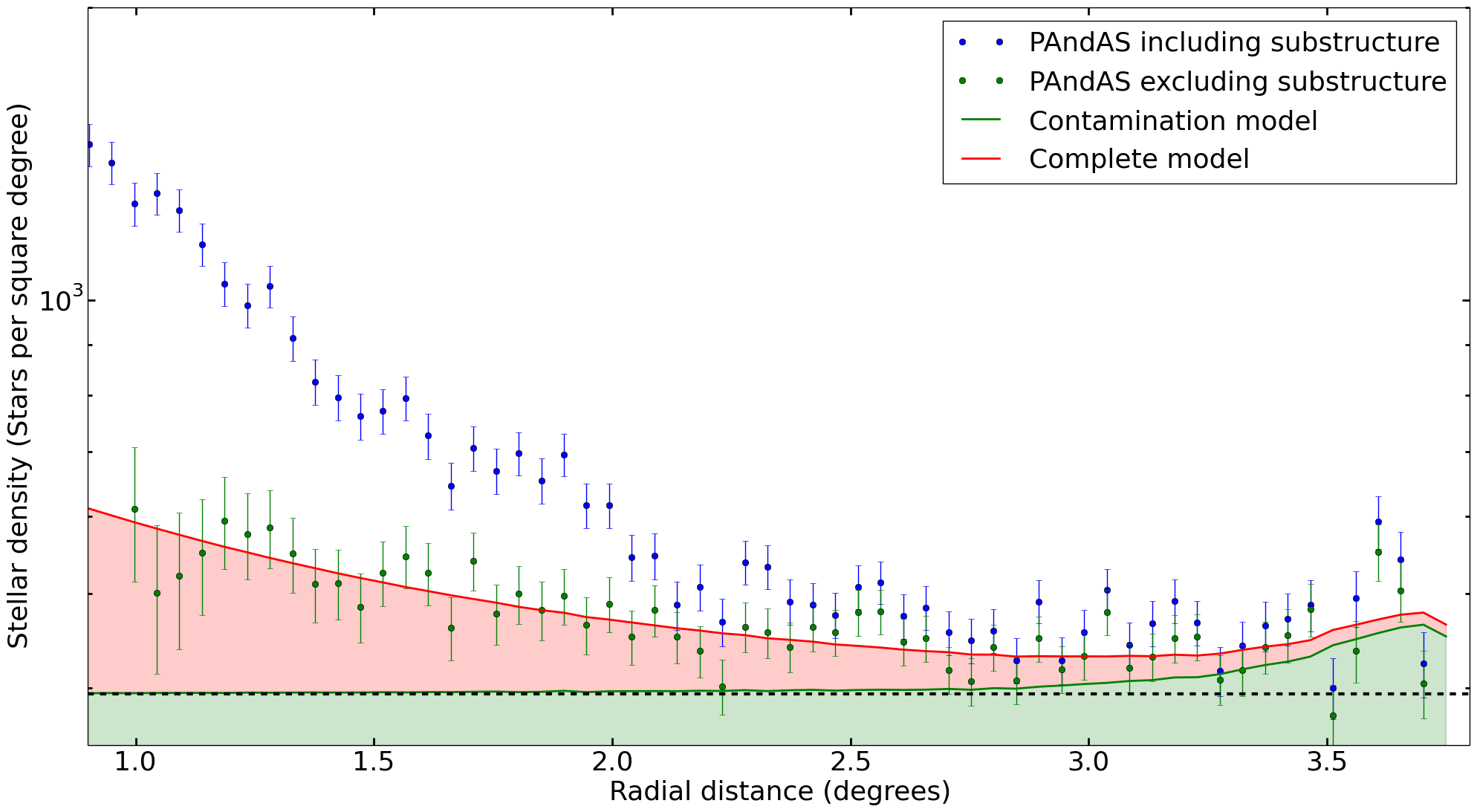}
\caption{Radially binned PAndAS data with $1\sigma$ error bars normalised by area. The upper solid (red) line marks the best radial fit, the lower solid (green) line marks the contribution of the contamination model.}
\label{fig:radial_cockcroft_m33} 
\end{figure}

\begin{figure}
\centering
\includegraphics[width=0.5\textwidth,clip=true]{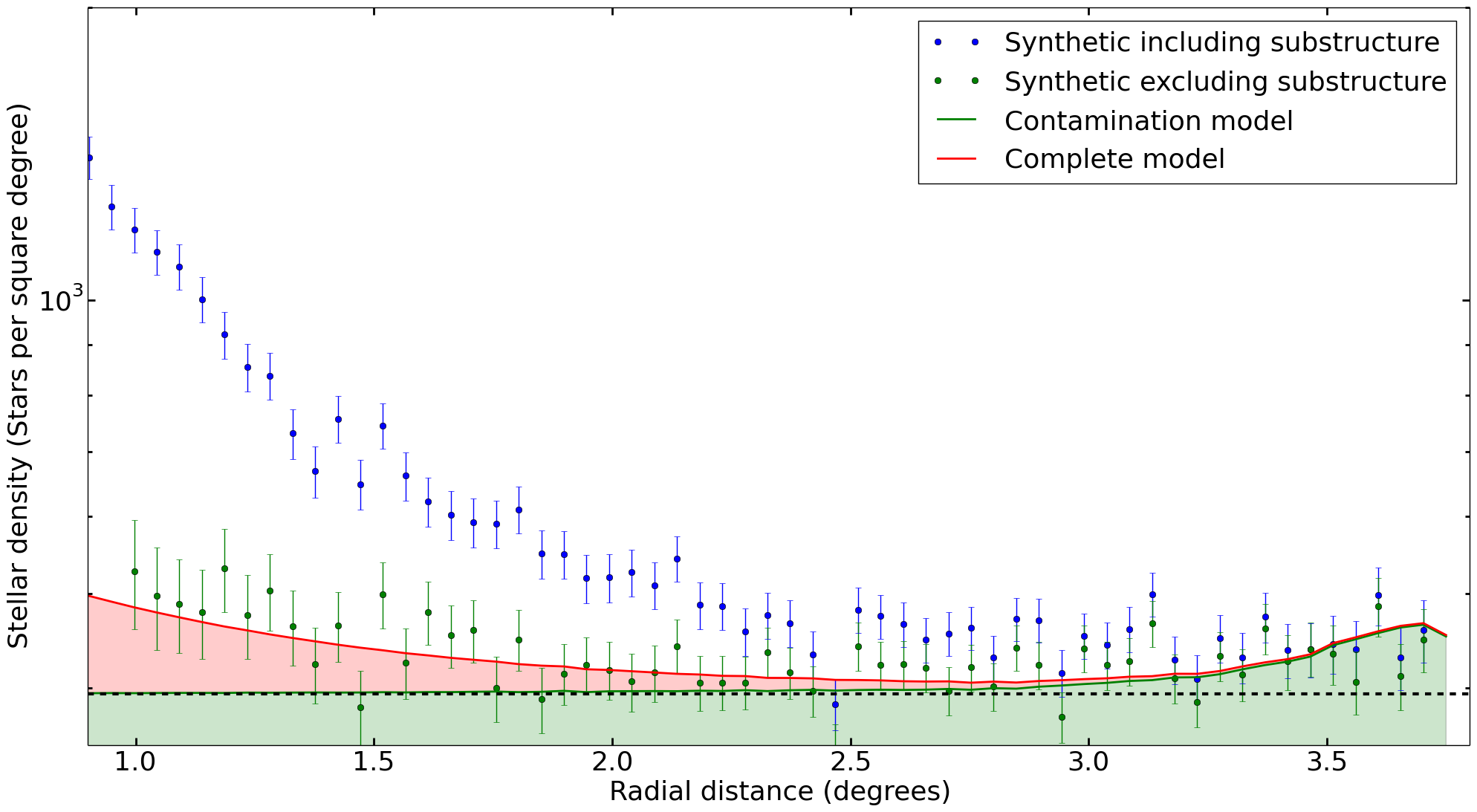}
\caption{Radially binned synthetic data with $1\sigma$ error bars normalised by area. The upper solid (red) line marks the best radial fit, the lower solid (green) line marks the contribution of the contamination model.}
\label{fig:radial_cockcroft_syn} 
\end{figure}

\begin{table*}
\begin{minipage}{120mm}
\caption{Cockcroft fits\label{tab:fit_cockcroft}}
\begin{tabular}{ l|c|c }
\hline
Dataset (binning, options) & \begin{tabular}{@{}c@{}}Halo central density \\ $\Sigma_0$ (stars per square degree)\end{tabular} & \begin{tabular}{@{}c@{}}Scale radius \\ $r_s$ (deg)\end{tabular} \\
\hline
Synthetic (ultra-fine, substructure, masked) & $385^{+512}_{-138} $ & $0.68^{+0.16}_{-0.17}$ \\ 
PAndAS (ultra-fine, masked) & $528^{+230}_{-102} $ & $1.01^{+0.14}_{-0.16}$ \\ 
\hline
\end{tabular}
\end{minipage}
\end{table*}

Following C13, we mask out the inner 1 degree and the regions within the grey contours are masked out, with the resulting area radially binned. Figures \ref{fig:radial_cockcroft_m33} and \ref{fig:radial_cockcroft_syn} show the radially binned data for the PAndAS data and synthetic data respectively, with the fitted parameters presented in Table \ref{tab:fit_cockcroft}. We find the signal of a smooth halo component in the PAndAS data, similar to what was found in C13. Intriguingly, we find a similar signal of a smooth halo component in the synthetic data, and, as no halo is present, this must represent a spurious detection due to the presence of substructure that cannot be removed via a simple magnitude-limit cut; this clearly brings into doubt the feature characterised in C13. The upturn visible in the contamination model at large radii is explained in more detail in Section \ref{sec:fit_radial}. What we have just shown is that this substructure can also masquerade as a halo when none is present. As the substructure is not well defined, and significantly brighter than the expected smooth halo, a signal for a smooth halo can be left behind by incorrectly masking or modelling the substructure.

With this result, and thus a loss of trust for this method, we test a range of other algorithms. For completeness, we begin with very simple algorithms, namely simple radial binning, and fits to the two-dimensional pixelated data. This is followed by a test of simple masking, in which we mask the coarsely pixelated data by simple density cuts. After this we test two novel algorithms, both of which seek to account for the substructure statistically. The first of these is the marginalised substructure fit, which treats the substructure as a nuisance parameter within each pixel. Our final method is the parametrised substructure fit, which allocates a free parameter for the substructure in each pixel. As noted earlier, these methods and the results of our tests with them are included in full in the Appendix. However, none of them are able to recover the parameters correctly when tested against the synthetic pixelations which include the centralised extended substructure. We find that the presence of substructure, like that which is found around M33, ruins all attempts to characterise an underlying smooth halo component.

However, the question of whether a stellar halo of M33 is present remains. As noted throughout this paper, substructure confuses all reasonable attempts to extract the properties of this component, but here we attempt to estimate just how bright a stellar halo could be hidden in the data. Based upon the various approaches presented in this paper, we estimate that, to be robustly detected, any halo component must have an average surface brightness of $\mu_g \sim 36$ mag arcsec$^{-2}$ ($\mu_i \sim 37$ mag arcsec$^{-2}$), over the range from 0.8 degrees to 3.75 degrees, which corresponds to a luminosity of approximately $10^6 L_\odot$ in the $V$-band. Extending this model to cover the entire M33 region from the centre out to 3.75 degrees, gives an average surface brightness of $\mu_g \sim 35$ mag arcsec$^{-2}$ ($\mu_i \sim 36$ mag arcsec$^{-2}$), roughly doubling the luminosity; further extending the model out to infinity does not significantly alter the luminosity. 

The total luminosity of M33 is estimated at approximately $10^9 L_\odot$ \citep{deVaucouleurs1991}. \citet{McConnachie2010} estimated the total luminosity of the extended substructure at approximately $10^7 L_\odot$, around one per cent of the luminosity of M33. In C13 the smooth halo component was limited to below $\sim 2.4 \times 10^6 L_\odot$ between 0.88 and 3.75 degrees from M33. Extrapolating their model inward raised this limit to $\sim 4 \times 10^6 L_\odot$, which was not significantly increased by expanding the model outward. We have estimated the luminosity at which our algorithms would definitely find the halo to be as low as $\sim 10^6 L_\odot$ over the region from 0.8 to 3.75 degrees, on the order of 0.1 per cent of the luminosity of M33 (or $\sim 2 \times 10^6 L_\odot$ extending the model inwards and outwards). But we expect the presence of substructure will continually frustrate its ultimate characterisation.

\section{Discussion and Conclusion}\label{sec:discussion}
The motivation of this research was to characterise the smooth stellar halo of M33, presented as a putative detection by C13; no claim to the origin of this component is made in this earlier work, and our characterisation will provide evidence to whether this component represents a halo formed from primordial and accreted components, or is in fact extended disk material, potentially distributed in the event that gave rise to the prominent gas/stellar warping of the disk of M33. Either result is significant for understanding galaxy evolution, with M33 sitting between the scale of the larger galaxies within the Local Group, which are known to host extensive stellar halos (e.g. \citealt{2005astro.ph..2366G,2014ApJ...787...30D}), and the Large Magellanic Cloud, in which an extensive stellar halo appears to be absent (\citealt{2010AJ....140.1719S}). Furthermore will provide  clues to the dynamical interactions in the history of the M31-M33 system (e.g. \citealt{McConnachie2009}).

Unlike previous approaches, this study employs an extant contamination model for various components, and uses robust statistical analyses to search for a signal of a smooth halo component. However, we have demonstrated a range of potential problems associated with detection of any putative halo of this spiral member of the Local Group. The model parameters are degenerate, the foreground contamination is structured and has not yet been completely characterised $-$ but principally, the signal for the halo is vanishingly faint, and completely degenerate with a significantly brighter extended substructure which pollutes the most desirable region around M33 to search for the halo. With such faint halo signatures as is expected for M33, the halo is dominated by every other component. Statistical fluctuations in any of these components can lead to significant changes to the fit of the halo. Thus even if the halo were detected, it would likely be inaccurately characterised, or with such large uncertainty bounds as to be unenlightening.

All of the fitting methods presented fail to detect any halo component in synthetic test data, which was designed to have a significantly brighter and more detectable halo component than any true halo component in the PAndAS data. These methods work on pixelation, and could be improved to avoid this (although removing pixelation from the parametrised substructure method would require significant changes), but this is unlikely to solve the key issue $-$ the PAndAS data alone is not sufficient to detect the halo of M33. 

The tentative detection of a possible, faint, extended stellar halo by C13, undertaken using PAndAS data with an inferior calibration and without the recent contamination model, describes an average surface brightness for the smooth halo component of less than $\mu_V=33$ magnitudes per square arc second. 
We find no evidence of a smooth halo component down to the limit reachable using the PAndAS data at approximately $\mu_V=35.5$. Using our methods, we are able to recover the halo parameters consistently down to ten times fainter than the halos used in our synthetic data tests. However, the presence of substructure such as is found around M33 will always complicate the fit as it cannot be separated from the signal using photometry alone, making a proper detection and characterisation of the halo impossible.

While we find no evidence of a smooth halo component around M33 in this work, there are indications in other studies. The detection of a handful of remote metal-poor globular clusters in the M33 system (\citealt{Stonkute2008}; \citealt{Huxor2009}; \citealt{Cockcroft2011}) provides evidence for the presence of a halo component. There has also been more direct evidence, with \citet{Chandar2002} identifying a kinematic signal of what could be a halo component around M33, and several metal poor RR Lyrae detections (\citealt{Sarajedini2006}; \citealt{Yang2010}; \citealt{Pritzl2011}). So the case remains open.

\section*{Acknowledgements}
BM acknowledges the support of an Australian Postgraduate Award. 
GFL thanks the Australian Research Council (ARC) for support through Discovery Project (DP110100678). GFL also gratefully acknowledges financial support through his ARC Future Fellowship (FT100100268). 
BJB was supported by a Marsden Fast-Start grant from the Royal Society of New Zealand. We thank the anonymous referee whose comments improved this paper.

\bibliography{Substructure}

\appendix
\section{Fitting Algorithms}

For completeness we present our suite of algorithms, and their results for both PAndAS data and synthetic mock data. All of the synthetic pixelations used here are generated from a model containing a smooth halo component defined by a central density $\Sigma_0$ of 800 stars per square degree, and a scale radius $r_s$ of 1.2 degrees (as discussed in Section \ref{sec:methods}).

\subsection{Simple Radial Fit}\label{sec:fit_radial}

\begin{table*}
\begin{minipage}{120mm}
\caption{Radial fits\label{tab:fit_radial}}
\begin{tabular}{ l|c|c }
\hline
Dataset (binning, options) & \begin{tabular}{@{}c@{}}Halo central density \\ $\Sigma_0$ (stars per square degree)\end{tabular} & \begin{tabular}{@{}c@{}}Scale radius \\ $r_s$ (deg)\end{tabular} \\
\hline
Synthetic (radial) & $742^{+122}_{-54}$ / $833^{+101}_{-60}$ & $1.24^{+0.13}_{-0.06}$ / $1.16^{+0.08}_{-0.05}$ \\
Synthetic (radial, substructure) & $4288^{+251}_{-183}$ & $0.68^{+0.02}_{-0.01}$ \\
PAndAS (radial) & $6080^{+427}_{-277}$ & $0.56^{+0.01}_{-0.01}$ \\
\hline
\end{tabular}
\medskip

Fits for two separate random realisations are given above for the synthetic pixelation (without included substructure) to show that the fit is able to correctly recover parameters, and is not biased above or below the true values.
\end{minipage}
\end{table*}

We start with the most basic fit, a simple radial binning. This is the only fitting method which does not directly use the pixelations discussed earlier, but rather a circular radial binning of the data. The synthetic data used is a re-binning of a finer grid than discussed earlier, to ensure high enough resolution for the re-binning. This binning still excludes stars within the M33 disk mask and beyond $3.75$~deg, and the model is normalised for these excluded regions in the respective radial bins. 
The likelihood function for this algorithm is the product of the Poisson probabilities of the data given the model in each bin, 
\begin{equation}\label{eq:radial_likelihood}
\mathcal{L} = \prod_{r=0.4}^{3.75} \frac{m_r^{d_r} \exp(-m_r)}{d_r!}
\end{equation}
where $d_r$ and $m_r$ are the data and model values respectively for each bin. 

When fitting the halo in the absence of substructure, this method is consistently able to recover the true values of the halo parameters within one sigma bounds (see example fits to random realisations of the synthetic data in Table \ref{tab:fit_radial}). However, once substructure is added, the fit fails robustly, with an estimate of the central halo density over $5$ times larger than the true value, and significantly underestimating the scale radius.

\begin{figure}
\centering
\includegraphics[width=0.5\textwidth,clip=true]{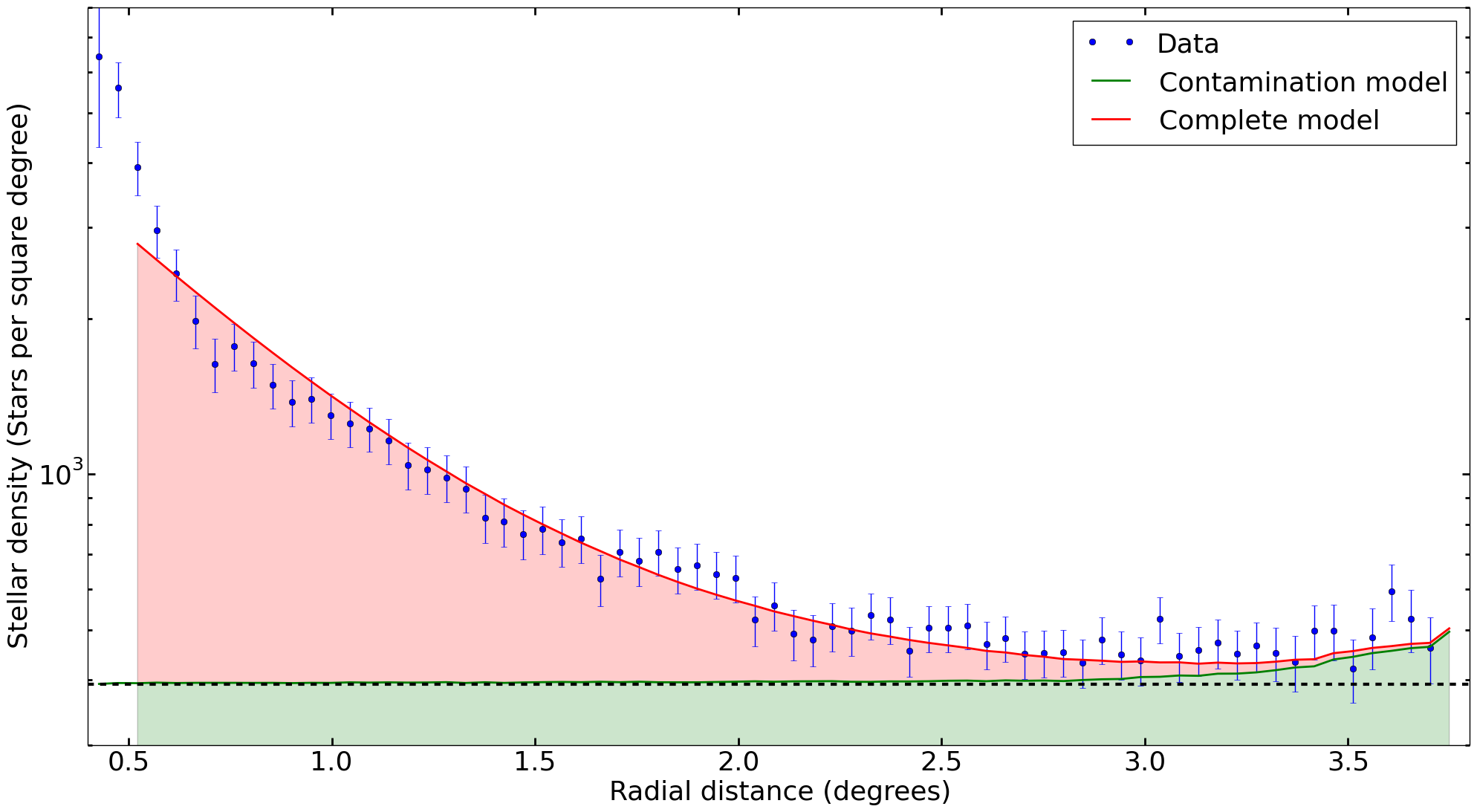}
\caption{Radially binned PAndAS data with $2\sigma$ error bars normalised by area. The upper solid (red) line marks the best radial fit, the lower solid (green) line marks the contribution of the contamination model, and the dashed (black) line marks a constant contamination level.}
\label{fig:radial_results} 
\end{figure}

Using this method on the PAndAS data produces deceptively promising results. Figure \ref{fig:radial_results} shows the radial fit to the data, with reasonable agreement, although there is clearly room for improvement. The fit fails at very small radii, as the substructure near the core increases rapidly, and fairly consistently overestimates the data at intermediate radii while underestimating the data at large radii. It is interesting to see the upturn of the contamination model at large radii, even when radially binning, producing a substantial departure from a constant contamination model. This upturn is due to the bias towards the North-East as the limits of the PAndAS footprint are reached in all other directions. Thus even using a radial fit, the new contamination model strongly influences the fit result, encouraging a shorter scale radius.

\begin{figure}
\centering
\includegraphics[width=0.5\textwidth,clip=true]{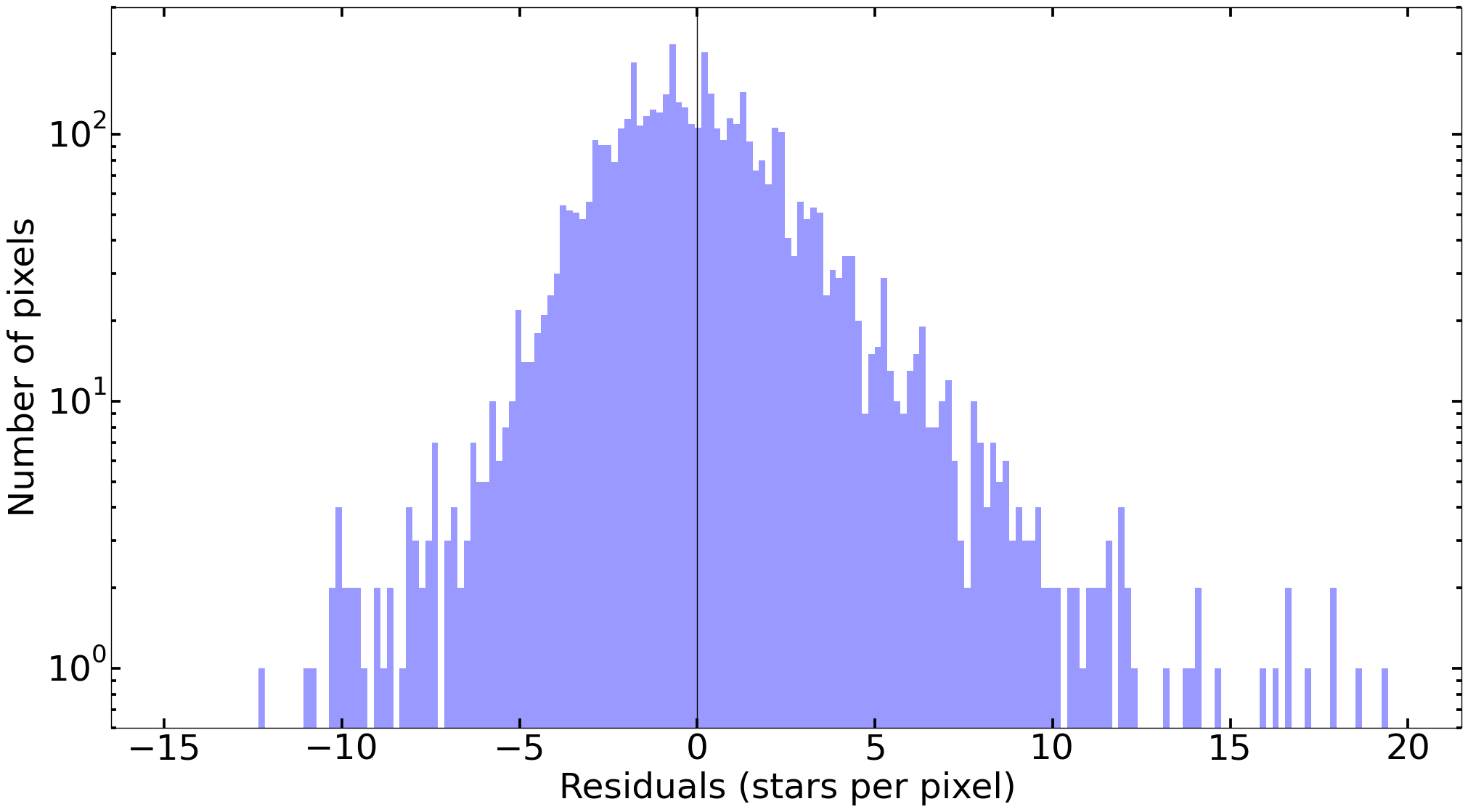}
\caption{Histogram of the residual pixel values in the coarse pixelation in Figure \ref{fig:m33_100} after subtracting the radial fit to the PAndAS data in Table \ref{tab:fit_radial}. The horizontal range excludes seven pixels with values between $21$ and $111$ stars per pixel. The vertical black line separates positive from negative.} 
\label{fig:radial_hist} 
\end{figure}

\begin{figure}
\centering
\includegraphics[width=0.5\textwidth,clip=true]{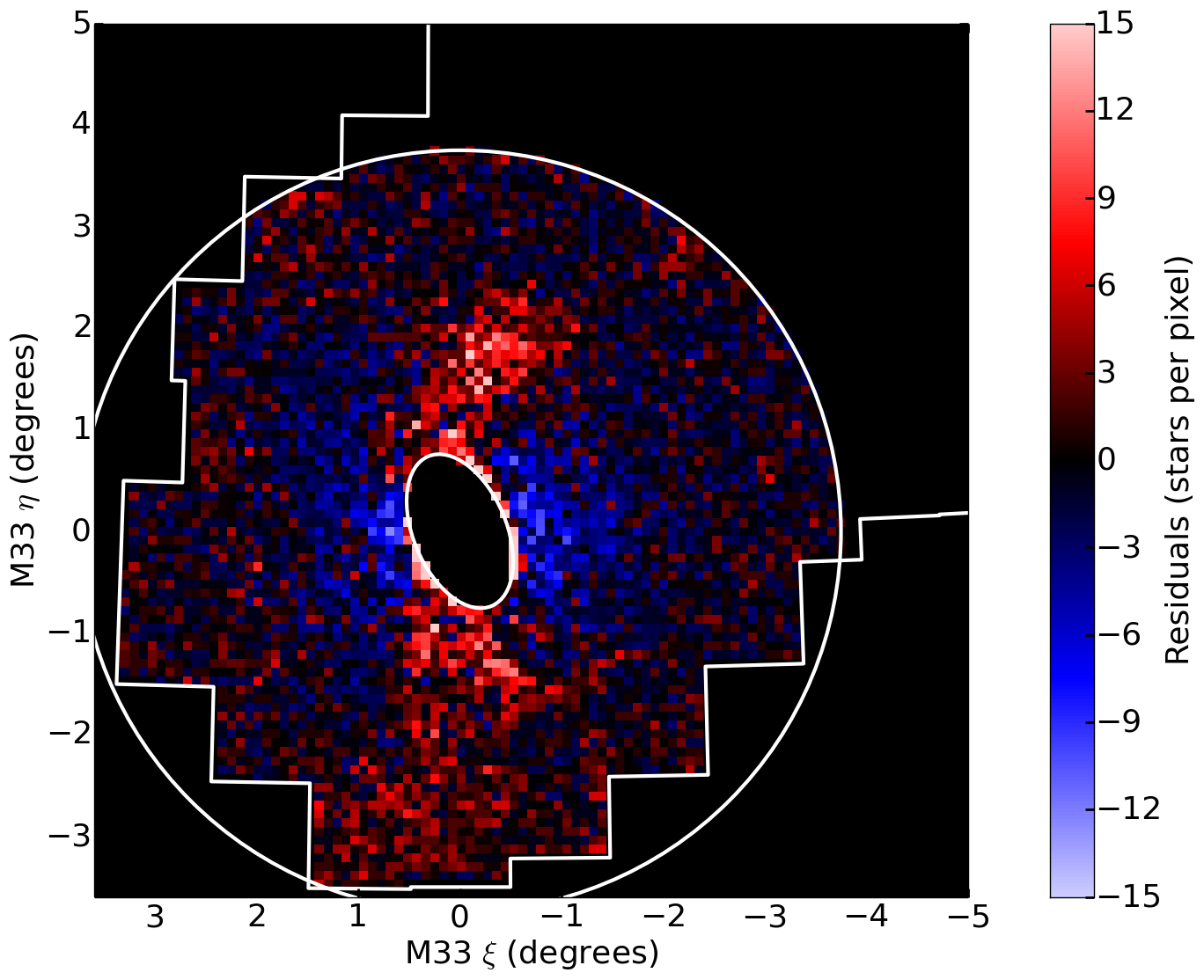}
\caption{Residual values from the coarse pixelation in Figure \ref{fig:m33_100} after subtracting the radial fit to the PAndAS data in Table \ref{tab:fit_radial}. The solid white circle marks the $3.75$~deg radius. The M33 disk mask is marked by a white ellipse, and the boundary of the PAndAS footprint is marked by a solid white line.}
\label{fig:radial_residuals} 
\end{figure}

Returning to the pixelations discussed earlier, Figure \ref{fig:radial_hist} shows a histogram of the residuals produced after subtracting the favoured model for the data. This is quite evenly split between positive and negative residuals, with a strong bias to small residuals and a slight bias to large positive residuals. While this is what would be expected from a good fit, by examining at the actual residual map in Figure \ref{fig:radial_residuals}, is it clear precisely how robustly the fit has failed. While the substructure is generally still positive, the halo has been significantly over-subtracted, particularly in the inner halo. From this it is apparent that a simple radial fit will not suffice.

\begin{table*}
\begin{minipage}{120mm}
\caption{Simple pixelation fits\label{tab:fit_pixelation}}
\begin{tabular}{ l|c|c }
\hline
Dataset (binning, options) & \begin{tabular}{@{}c@{}}Halo central density \\ $\Sigma_0$ (stars per square degree)\end{tabular} & \begin{tabular}{@{}c@{}}Scale radius \\ $r_s$ (deg)\end{tabular} \\
\hline
Synthetic (coarse) & $780^{+55}_{-52}$ / $821^{+59}_{-53}$ & $1.22^{+0.06}_{-0.06}$ / $1.16^{+0.06}_{-0.06}$ \\
Synthetic (fine) & $797^{+63}_{-51}$ / $753^{+56}_{-49}$ & $1.19^{+0.06}_{-0.06}$ / $1.24^{+0.06}_{-0.06}$ \\
Synthetic (coarse, substructure) & $4483^{+170}_{-164}$ & $0.67^{+0.01}_{-0.01}$ \\
Synthetic (fine, substructure) & $4548^{+165}_{-170}$ & $0.67^{+0.01}_{-0.01}$ \\
\hline
\end{tabular}
\medskip

As for Table \ref{tab:fit_radial}, fits for two separate random realisations are given above for the synthetic pixelations (without included substructure) to show that the fit is able to correctly recover parameters, and is not biased above or below the true values.
\end{minipage}
\end{table*}

\subsection{Simple Pixelation Fit}\label{sec:fit_pixelation}
Returning to two-dimensional binning enables a more spatially resolved analysis, and allows us to take full advantage of the contamination model. The likelihood function is slightly modified from Equation \ref{eq:radial_likelihood} to
\begin{equation}\label{eq:likelihood_pixelated}
\mathcal{L} = \prod \frac{m_{(\xi,\eta)}^{d_{(\xi,\eta)}} \exp(-m_{(\xi,\eta)})}{d_{(\xi,\eta)}!}
\end{equation}
where $d_{(\xi,\eta)}$ and $m_{(\xi,\eta)}$ are the data and model values respectively for each usable pixel. 

As with the radial fit, we start by ensuring that this method can reliably recover the parameters of a synthetic dataset in the absence of substructure. Fits to example random realisations are given in Table \ref{tab:fit_pixelation}. This method reliably recovers the parameter values within two sigma uncertainty bounds, and generally within one sigma. The uncertainty bounds are roughly the same size for coarse and fine pixelations, and are both smaller than for the results of radial fitting method on corresponding datasets. As with the radial method, adding substructure results in a complete failure to recover the true parameter values. It is not possible to fit the stellar halo without dealing with the substructure.

\subsection{Masked Substructure Fit}\label{sec:fit_masked}

\begin{table*}
\begin{minipage}{120mm}
\caption{Masked substructure fits\label{tab:fit_masked}}
\begin{tabular}{ l|c|c }
\hline
Dataset (binning, options) & \begin{tabular}{@{}c@{}}Halo central density \\ $\Sigma_0$ (stars per square degree)\end{tabular} & \begin{tabular}{@{}c@{}}Scale radius \\ $r_s$ (deg)\end{tabular} \\
\hline
Synthetic (coarse, substructure, mask$>10$) & $1161^{+106}_{-99} $ & $1.00^{+0.05}_{-0.05}$ \\
Synthetic (coarse, substructure, mask$>9$) & $994^{+111}_{-104}$ & $0.99^{+0.06}_{-0.06}$ \\
Synthetic (coarse, substructure, mask$>8$) & $753^{+120}_{-98}$ & $1.02^{+0.08}_{-0.08}$ \\
Synthetic (coarse, substructure, mask$>7$) & $612^{+125}_{-109}$ & $0.94^{+0.10}_{-0.09}$ \\
\hline
\end{tabular}
\end{minipage}
\end{table*}

\begin{figure}
\centering
\includegraphics[width=0.5\textwidth,clip=true]{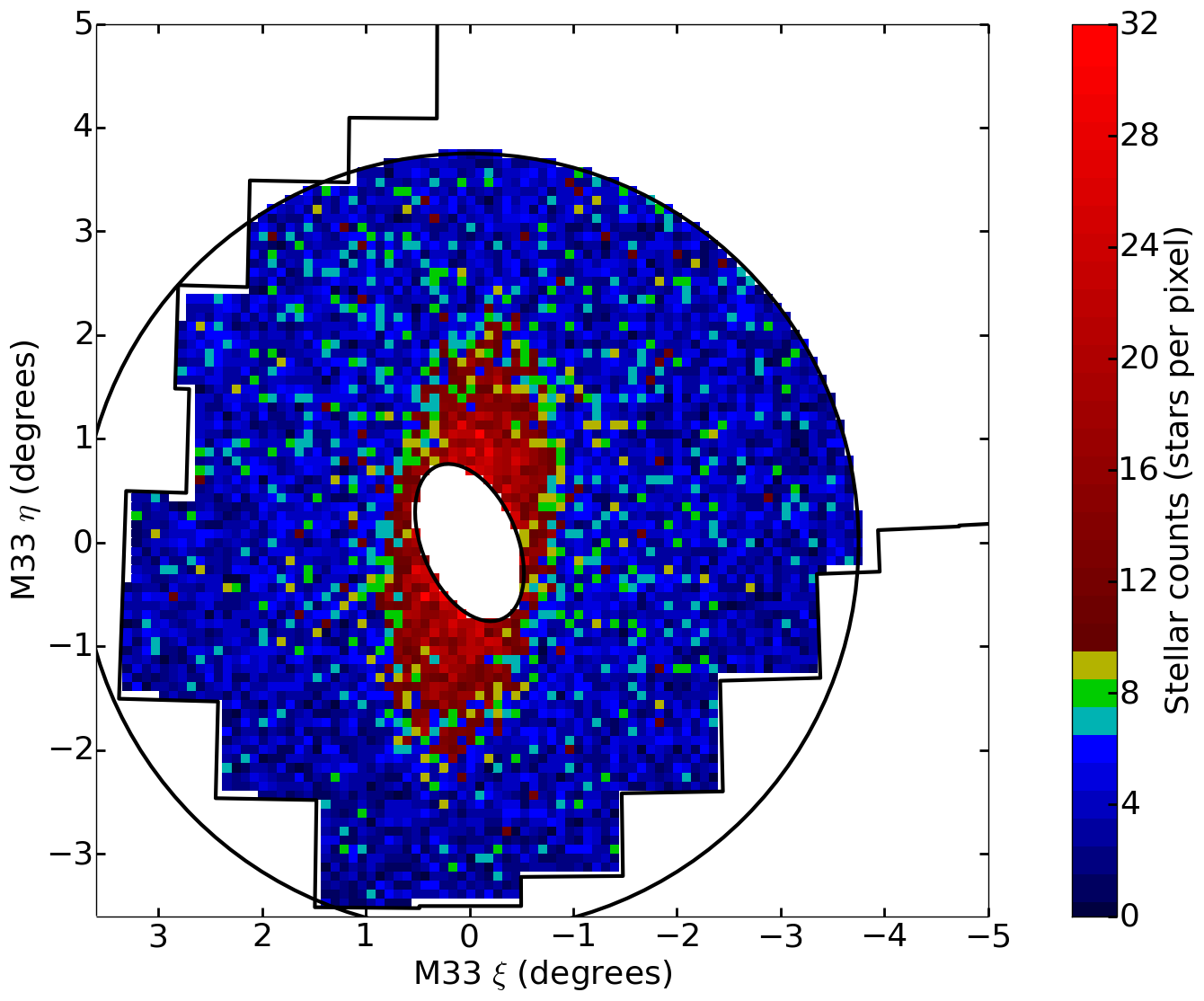}
\caption{Random realisation of the synthetic data, including central substructure, designed to match the coarse pixelation of the stellar data in Figure \ref{fig:m33_100}. The colour-axis has been selected to highlight the different cuts used in this section. The solid black circle marks the $3.75$~deg radius. The M33 disk mask is marked by a black ellipse, and the boundary of the PAndAS footprint is marked by a solid black line.}
\label{fig:syn_cuts} 
\end{figure}

It is common to resort to masking to fit a faint signal in the presence of a another dominant signal. We believe this approach is suboptimal, and masking should generally be avoided. We attempt masking here for completeness and to demonstrate its failings. This method uses the same likelihood function as the previous section, and simply excludes further pixels. By removing the pixels with the largest number of stars in them, we expect to selectively exclude the denser substructure which is obfuscating the diffuse halo signal. 

The data in the fine pixelation is too spread out to use this method, since the majority of pixels contain $\leq1$ star. Thus, for this method we only use the coarse pixelation. We test four different cuts, masking down to successively fainter levels each time. Figure \ref{fig:syn_cuts} shows the various cuts tested, with red pixels excluded by all masks, gold first excluded by the second mask, green first excluded by the third mask, teal only excluded by the final mask, and blue not excluded by any mask. The results of the fits are presented in Table \ref{tab:fit_masked}, with all fits performed on the same random realisation of synthetic data.

There appears to be a large initial improvement from the fit to the unmasked coarse synthetic data, with each subsequent cut removing more pixels and closing in on the true parameter values $-$ but this is not the case. As further pixels are masked out, the fit to $\Sigma_0$ reaches and then passes the true value. Selecting the value at which the mask excludes pixels, enables us to control the value which the fit returns for the central halo density, invalidating the results completely. Furthermore, this method only successfully recovers the value for the scale radius by the virtue of the uncertainty range increasing as the masks remove more pixels. Finally, the uncertainty range for the central density is unsatisfactorily large, at double the size of the substructure-free fits in the previous section.

It should not be surprising that this method is unsatisfactory, as it discards a significant portion of the data. It is also unlikely that masking can remove such dominant contamination properly, as it will bleed into nearby unmasked pixels, and due to its dominance, missing even a small portion of the contamination has a large effect on the result of the fit.

In this case in particular, it is unclear what criteria could be used to mask out a dominant contamination signal which lies in the same place in parameter space, especially when the target signal is expected to be most dense in the same regions that the contamination is strongest. A more statistical approach is needed to solve this.

\subsection{Marginalised Substructure Fit}\label{sec:fit_marginalised}

\begin{table*}
\begin{minipage}{120mm}
\caption{Marginalised substructure fits\label{tab:fit_marginalised}}
\begin{tabular}{ l|c|c }
\hline
Dataset (binning, options) & \begin{tabular}{@{}c@{}}Halo central density \\ $\Sigma_0$ (stars per square degree)\end{tabular} & \begin{tabular}{@{}c@{}}Scale radius \\ $r_s$ (deg)\end{tabular} \\
\hline
Synthetic (coarse, $\alpha=1$) & $1363^{+220}_{-193} $ & $0.51^{+0.04}_{-0.04}$ \\
Synthetic (coarse) & $815^{+72}_{-59} $ & $1.14^{+0.07}_{-0.07}$ \\
Synthetic (coarse, substructure) & $6302^{+364}_{-344} $ & $0.51^{+0.02}_{-0.02}$ \\
\hline
\end{tabular}
\end{minipage}
\end{table*}

This method was developed specifically to address the problems in the masked substructure fitting (Section~\ref{sec:fit_masked}). Instead of ignoring or removing the substructure, we marginalise over the substructure within each pixel as a nuisance parameter. We expect the substructure to be reasonably well represented by an exponential distribution, as there should be many small values and only a few large values, thus we include an exponential prior on the substructure, $P(s_{(\xi,\eta)}) = \alpha \exp\left(-\alpha s_{(\xi,\eta)}\right)$, where $\alpha$ is inverse of the mean of the exponential distribution and assuming no pixel-to-pixel correlations. We still use a Poisson probability for the value of the data within each pixel, and thus by marginalising over all possible values for the substructure component, the likelihood becomes  
\begin{equation}
\begin{split}
\mathcal{L} & = \prod \int_{0}^{\infty} P\left(d_{(\xi,\eta)}|m_{(\xi,\eta)}+s_{(\xi,\eta)}\right)P\left(s_{(\xi,\eta)}\right)ds_{(\xi,\eta)} \\
& \\
&= \prod \int_{0}^{\infty} ds_{(\xi,\eta)} \left(\alpha \exp\left(-\alpha s_{(\xi,\eta)}\right)\right)  \\
&\ \ \ \ \times \left(\frac{\left(m_{(\xi,\eta)}+s_{(\xi,\eta)}\right)^{d_{(\xi,\eta)}} \exp\left(-\left(m_{(\xi,\eta)}+s_{(\xi,\eta)}\right)\right)}{d_{(\xi,\eta)}!}\right)  \\
& \\
&= \prod \frac{\alpha \exp\left(\alpha m_{(\xi,\eta)}\right) \Gamma\left(d_{(\xi,\eta)}+1, m_{(\xi,\eta)}(1+\alpha)\right)}{\left(1+\alpha\right)^{\left(d_{(\xi,\eta)}+1\right)}\Gamma\left(d_{(\xi,\eta)}+1\right)}
\end{split}
\label{eq:likelihood_marginalised}
\end{equation}
where $d_{(\xi,\eta)}$, $m_{(\xi,\eta)}$, and $s_{(\xi,\eta)}$ are the data, model, and unknown substructure values respectively for each usable pixel. $\Gamma(x,y)$ and $\Gamma(x)$ are the incomplete and complete Gamma functions respectively.

The value of the $\alpha$ parameter is dependant on the particular substructure present within the data. When $\alpha$ is large, the likelihood returns to a simple Poisson probability, as in the simple pixelation method, and is the optimal choice in the absence of substructure. 
This parameter should be left free for the algorithm to optimise based on the data, and attempting to set the value for $\alpha$ results in a sub-optimal fit. The two substructure-free fits in Table \ref{tab:fit_marginalised} are both for the same random realisation of synthetic data, however the fit with $\alpha$ set to one gives a poor fit, while the other fit recovers the parameters within one sigma.

To boost the signal of the substructure within each pixel, the coarse pixelations were used in favour of the fine pixelations. 
This method consistently recovers the parameters for substructure-free synthetic data within two sigma, and generally within one sigma. It performs slightly worse than the simple pixelation method, although with comparable uncertainty bounds, as it is designed to operate in the presence of substructure. Unfortunately, this method is still unable to cope with the intense central substructure degenerate with the halo signal, as can be seen by the failed fit in Table \ref{tab:fit_marginalised}.

\begin{table*}
\begin{minipage}{120mm}
\caption{Parametrised substructure fits\label{tab:fit_parametrised}}
\begin{tabular}{ l|c|c }
\hline
Dataset (binning, options) & \begin{tabular}{@{}c@{}}Halo central density \\ $\Sigma_0$ (stars per square degree)\end{tabular} & \begin{tabular}{@{}c@{}}Scale radius \\ $r_s$ (deg)\end{tabular} \\
\hline
Synthetic (coarse, substructure) & $4820^{+1363}_{-1430} $ & $0.46^{+0.09}_{-0.06}$ \\ 
PAndAS (coarse) & $8964^{+1895}_{-1816}$ & $0.36^{+0.03}_{-0.03}$ \\ 
\hline
\end{tabular}
\end{minipage}
\end{table*}
	
\subsection{Parametrised Substructure Fit}\label{sec:fit_parametrised}
We have shown that the substructure cannot be ignored or masked, and that even marginalisation is not sufficient to detect any underlying halo component. This last method models the substructure distribution directly. A free parameter $s_{(\xi,\eta)}$ is assigned to each pixel, to represent the substructure present. The pixelation of the substructure is then smoothed using a Gaussian with a dispersion of $\sigma = 0.087\degr$, or one pixel width. This smoothing speeds up the fit, and also performs a simple regularization, as it is expected that substructure would be present in pixels in close proximity to other substructure containing pixels. 

On each step, the non-substructure parameters are stepped in the standard Metropolis-Hastings fashion, and the substructure parameters are each stepped with a probability of $0.5$ per cent. This ensured that only a small change to the substructure would be made each time, resulting in a larger portion of steps being accepted. Each substructure parameter stepped is given a new value from the distribution 

\begin{equation}\label{eq:substructure}
F(\alpha) = 
  \begin{cases}
    0,       & \quad \text{if } 0<\alpha\leq\beta \\
    -\mu \ln\left(1-\frac{\alpha-\beta}{1-\beta}\right), & \quad \text{if } \beta\leq\alpha<1\\
  \end{cases}
\end{equation}
based on a random value of $\alpha$ between $0$ and $1$, where $\beta$ is the proportion of pixels expected not to contain any substructure, and $\mu$ is the mean of the exponential distribution representing the substructure in the remaining pixels. This distribution function acts as a prior on the substructure, without which the algorithm would determine the data to be perfectly represented by substructure alone. The initial values for all substructure parameters are zero.
$\beta$ and $\mu$ are optimised by the fitting algorithm, as setting these parameters manually results in a sub-optimal fit by defining the substructure present in the data, as with $\alpha$ in the marginalised substructure fitting method.

Due to the complexity of this algorithm, the affine-invariant ensemble sampler was discarded and replaced with a standard Metropolis-Hastings sampler. This slowed the fit further, especially considering the degeneracy between the halo parameters, and the large increase in parameters overall. However, the likelihood function is now reduced to a simple Poisson probability,
\begin{equation}\label{eq:likelihood_parametrised}
\mathcal{L} = \prod \frac{\left(m_{(\xi,\eta)}+s_{(\xi,\eta)}\right)^{d_{(\xi,\eta)}} \exp\left(-\left(m_{(\xi,\eta)}+s_{(\xi,\eta)}\right)\right)}{d_{(\xi,\eta)}!}  \\
\end{equation}
where $d_{(\xi,\eta)}$, $m_{(\xi,\eta)}$, and $s_{(\xi,\eta)}$ are the data, model, and smoothed substructure values respectively for each usable pixel.

\begin{figure}
\centering
\includegraphics[width=0.5\textwidth,clip=true]{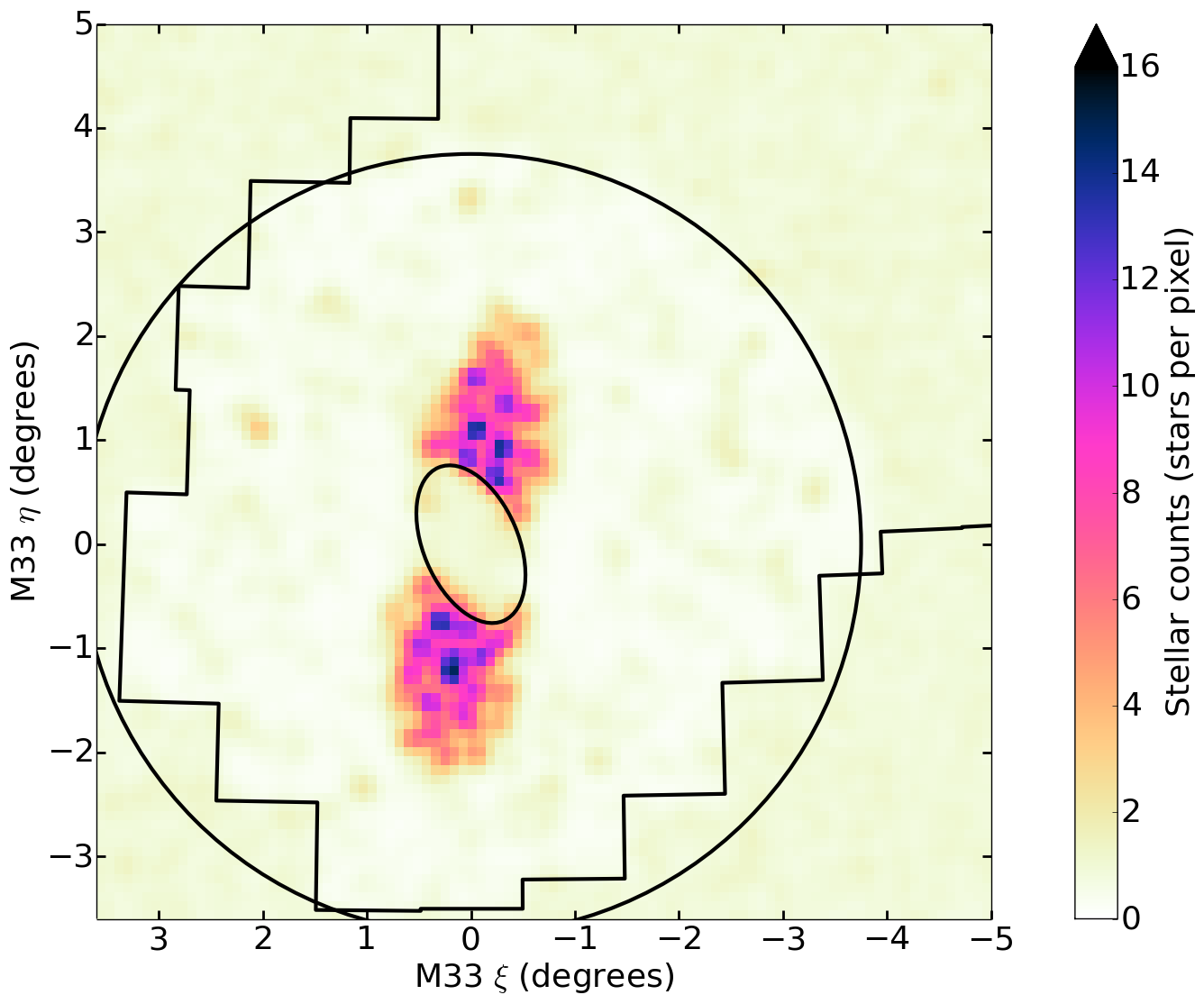}
\caption{Substructure model generated by the fit to a random realisation of coarse synthetic data with central substructure by the parametrised substructure method. The fit parameters are given in Table \ref{tab:fit_parametrised}. The solid black circle marks the $3.75$~deg radius. The M33 disk mask is marked by a black ellipse, and the boundary of the PAndAS footprint is marked by a solid black line. } 
\label{fig:param_syn_long_model} 
\end{figure}
\begin{figure}
\centering 
\includegraphics[width=0.5\textwidth,clip=true]{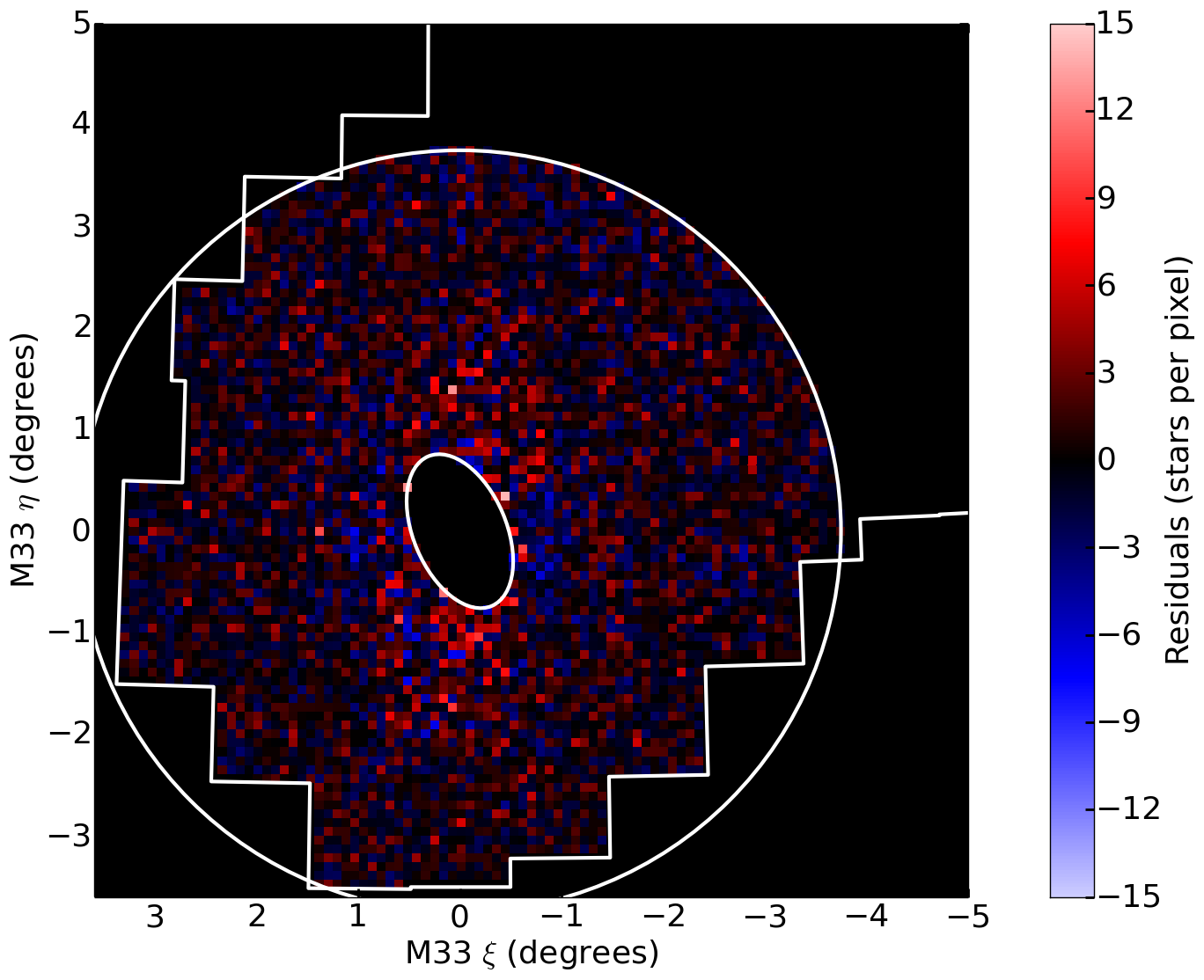}
\caption{Residual map for the fit to a random realisation of coarse synthetic data with central substructure by the parametrised substructure method. The residuals for this fit range from $-13$ to $15$. The fit parameters are given in Table \ref{tab:fit_parametrised}. The solid white circle marks the $3.75$~deg radius. The M33 disk mask is marked by a white ellipse, and the boundary of the PAndAS footprint is marked by a solid white line.} 
\label{fig:param_syn_long_residuals} 
\end{figure}

To boost the signal of the substructure within each pixel, and also to reduce the run-time of the algorithm, the coarse pixelations were used in favour of the fine pixelations. The fits to the synthetic data with substructure did not recover the halo parameters, as seen by the example in Table \ref{tab:fit_parametrised}. The substructure is reasonably well reproduced away from the M33 disk, as shown in Figure \ref{fig:param_syn_long_model}, however, three overdensities are recovered in the North-East quadrant which are only statistical fluctuations in the background. This should serve as a warning to over-interpreting the model recovered by this method. 

The residuals based on this fit, shown in Figure \ref{fig:param_syn_long_residuals}, are surprisingly small. This indicates that there is a further degeneracy plaguing the fits. The fitting algorithm is pruning the central substructure until it fits the circular model expected for the halo, and then fitting a halo to this. As was already discussed, this substructure is indistinguishable from the halo signal using this data alone, and thus this method is unable to recover the halo.

\begin{figure}
\centering
\includegraphics[width=0.5\textwidth,clip=true]{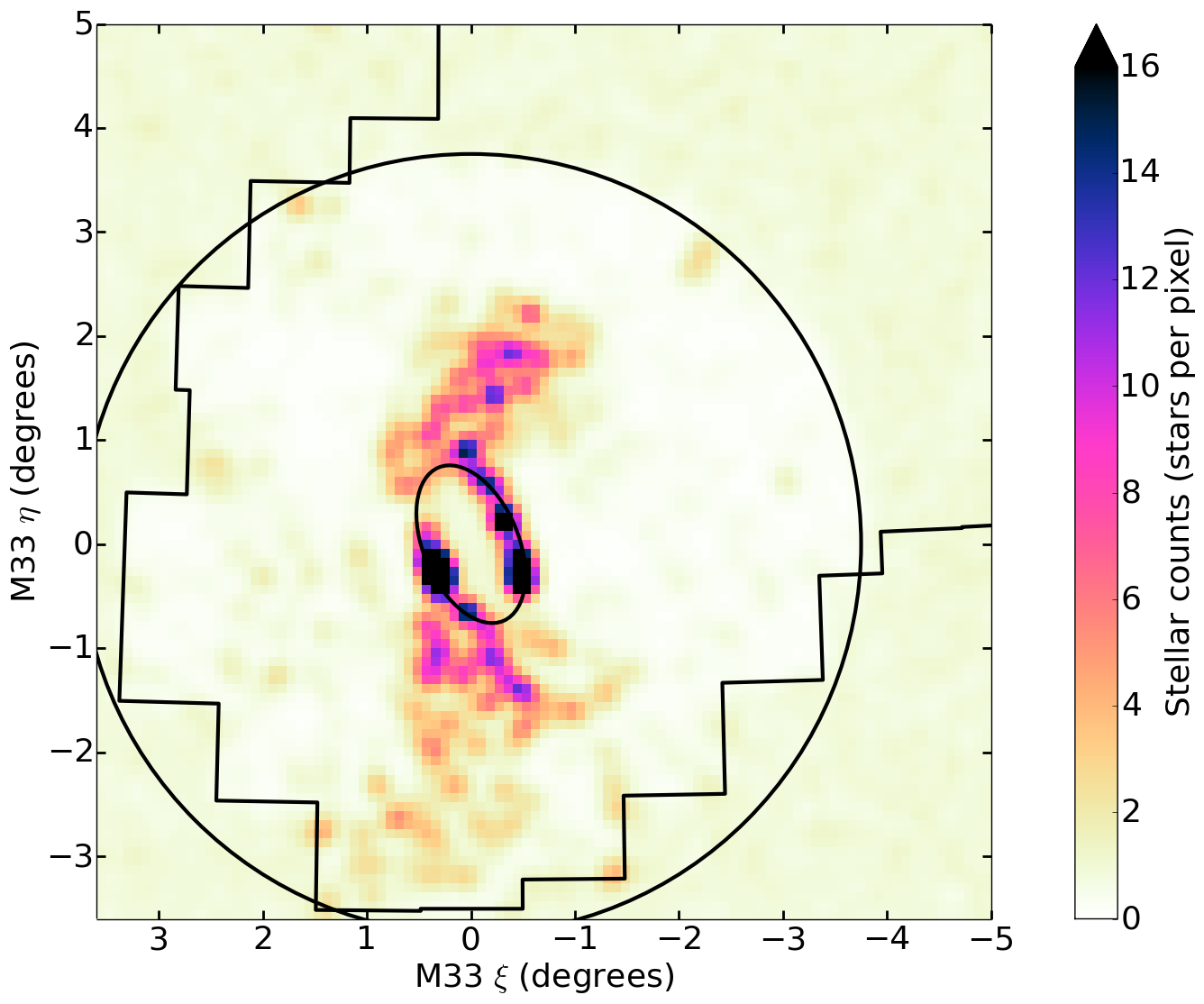}
\caption{The substructure model generated by sampling the fit to the coarse pixelation of the PAndAS data by the parametrised substructure method. The fit parameters are given in Table \ref{tab:fit_parametrised}. The solid black circle marks the $3.75$~deg radius. The M33 disk mask is marked by a black ellipse, and the boundary of the PAndAS footprint is marked by a solid black line. } 
\label{fig:param_data_long_model} 
\end{figure}
\begin{figure} 
\centering 
\includegraphics[width=0.5\textwidth,clip=true]{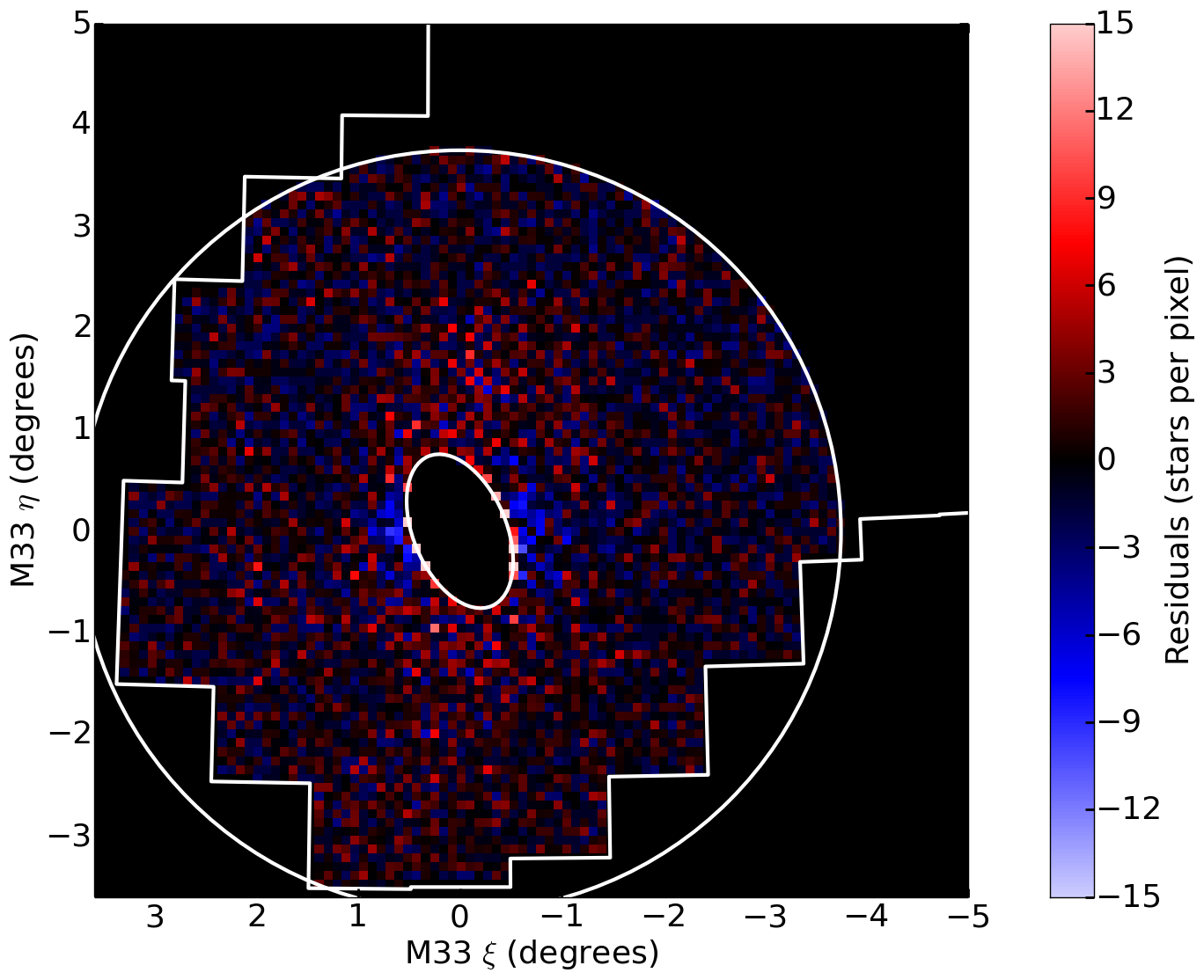}
\caption{Residual map for the fit to the coarse pixelation of the PAndAS data by the parametrised substructure method. The residuals for this fit range from $-10$ to $41$. The fit parameters are given in Table \ref{tab:fit_parametrised}. The solid white circle marks the $3.75$~deg radius. The M33 disk mask is marked by a white ellipse, and the boundary of the PAndAS footprint is marked by a solid white line.} 
\label{fig:param_data_long_residuals} 
\end{figure}

While the fit for the halo parameters is poor, the substructure model is reasonably reliable away from the M33 core. Thus this method can be used to recover a rough model of the substructure around M33, shown in Figure \ref{fig:param_data_long_model}. Within the $3.75$~deg cut-off, all visible substructure has been recovered. This includes the globular cluster systems in the North-West, a faint over-density at ($1,3$), a faint over-density at ($2,-1$ to $1$), a larger over-density to the South, and the rough structure of the central substructure to the North and South. 
The residual map of this fit in Figure \ref{fig:param_data_long_residuals} shows the failure of the fit to the halo, with large positive and negative residuals at small radii.

\label{lastpage}

\end{document}